\documentclass[twocolumn]{aastex631}

\usepackage{amsmath}
\usepackage{threeparttable}

\begin{document}

\title{Investigating the Star-Formation Characteristics of Radio Active Galactic Nuclei}

\author[0009-0009-8209-4613]{$\text{Bojun Zhang}^{\star}$}

\affiliation{Department of Astronomy and Astrophysics,
525 Davey Lab,
The Pennsylvania State University,
University Park,
PA 16802, USA}
\affiliation{CAS Key Laboratory for Research in Galaxies and Cosmology, Department of Astronomy, University of Science and Technology of China, Hefei 230026, China}
\affiliation{School of Astronomy and Space Sciences,
University of Science and Technology of China,
Hefei 230026,
China}
\author[0000-0002-4436-6923]{$\text{Fan Zou}^{\bigtriangleup}$}
\affiliation{Department of Astronomy, University of Michigan, 1085 S University, Ann Arbor, MI 48109, USA}
\affiliation{Department of Astronomy and Astrophysics,
525 Davey Lab,
The Pennsylvania State University,
University Park,
PA 16802, USA}
\affiliation{Institute for Gravitation and the Cosmos, The Pennsylvania State University, University Park, PA 16802, USA}
\author[0000-0002-0167-2453]{W. N. Brandt}
\affiliation{Department of Astronomy and Astrophysics,
525 Davey Lab,
The Pennsylvania State University,
University Park,
PA 16802, USA}
\affiliation{Institute for Gravitation and the Cosmos, The Pennsylvania State University, University Park, PA 16802, USA}
\affiliation{Department of Physics, 104 Davey Laboratory, The Pennsylvania State University, University Park, PA 16802, USA}

\author[0000-0002-1653-4969]{Shifu Zhu}
\affiliation{CAS Key Laboratory for Research in Galaxies and Cosmology, Department of Astronomy, University of Science and Technology of China, Hefei 230026, China}
\affiliation{School of Astronomy and Space Sciences,
University of Science and Technology of China,
Hefei 230026,
China}

\author[0000-0001-6317-8488]{Nathan Cristello}
\affiliation{Department of Astronomy and Astrophysics,
525 Davey Lab,
The Pennsylvania State University,
University Park,
PA 16802, USA}
\author[0000-0002-8577-2717]{Qingling Ni}
\affiliation{Max-Planck-Institut f\"{u}r extraterrestrische Physik (MPE), Gie{\ss}enbachstra{\ss}e 1, D-85748 Garching bei M\"unchen, Germany}

\author[0000-0002-1935-8104]{Yongquan Xue}
\affiliation{CAS Key Laboratory for Research in Galaxies and Cosmology, Department of Astronomy, University of Science and Technology of China, Hefei 230026, China}
\affiliation{School of Astronomy and Space Sciences,
University of Science and Technology of China,
Hefei 230026,
China}

\author[0000-0002-6990-9058]{Zhibo Yu}
\affiliation{Department of Astronomy and Astrophysics,
525 Davey Lab,
The Pennsylvania State University,
University Park,
PA 16802, USA}
\affiliation{Institute for Gravitation and the Cosmos, The Pennsylvania State University, University Park, PA 16802, USA}

\email{$^{\star}$bjz5234@psu.edu}
\email{$^{\bigtriangleup}$fanzou01@gmail.com}


\begin{abstract}
The coevolution of supermassive black holes and their host galaxies represents a fundamental question in astrophysics. One approach to investigating this question involves comparing the star-formation rates (SFRs) of active galactic nuclei (AGNs) with those of typical star-forming galaxies. At relatively low redshifts ($z\lesssim 1$), radio AGNs manifest diminished SFRs, indicating suppressed star formation, but their behavior at higher redshifts is unclear. To examine this, we leveraged galaxy and radio AGN data from the well-characterized W-CDF-S, ELAIS-S1, and XMM-LSS fields. We established two mass-complete reference star-forming galaxy samples and two radio AGN samples, consisting of 1,763 and 6,766 radio AGNs, the former being higher in purity and the latter more complete. We subsequently computed star-forming fractions ($f_{\text{SF}}$; the fraction of star-forming galaxies to all galaxies) for galaxies and radio-AGN-host galaxies and conducted a robust comparison between them up to $z\approx3$. We found that the tendency for radio AGNs to reside in massive galaxies primarily accounts for their low $f_{\text{SF}}$, which also shows a strong negative dependence upon $M_{\star}$ and a strong positive evolution with $z$. To investigate further the star-formation characteristics of those star-forming radio AGNs, we constructed the star-forming main sequence (MS) and investigated the behavior of the position of AGNs relative to the MS at $z\approx0-3$. Our results reveal that radio AGNs display lower SFRs than star-forming galaxies in the low-$z$ and high-$M_{\star}$ regime and, conversely, exhibit comparable or higher SFRs than MS star-forming galaxies at higher redshifts or lower $M_{\star}$. 
\end{abstract}
                     



\section{Introduction} \label{sec: intro}


Active galactic nuclei (AGNs) are believed to play a crucial role in the evolution of their host galaxies through the process of feedback, which can regulate or even quench star formation (e.g., \citealt{Kormendy&Ho+13}; \citealt{Weinburger+18}).  
In particular, radio AGNs, which are characterized by powerful relativistic jets, can potentially heat the surrounding gas and prevent it from collapsing to form new stars, a process known as ``radio-mode" feedback (e.g., \citealt{Hardcastle+19}; \citealt{Kondapally+22}). Observational evidence for this feedback has been found both in the nearby universe  (e.g., \citealt{Alatalo+15}; \citealt{Salome+16}; \citealt{Jarvis+21}; \citealt{Drevet+23}) and at higher redshifts (e.g., \citealt{Maiolino+12}). However, theoretical studies have shown radio jets may also trigger shock waves that compress gas and induce star formation in some cases (e.g., \citealt{Silk+13}; \citealt{Bieri+16}; \citealt{Fragile+17}; \citealt{Mukherjee+18}). Direct observational evidence for this positive feedback has been found in rare cases (e.g., \citealt{Salome+15}; \citealt{Lacy+17}), highlighting the complex interplay between radio AGN activity and star formation processes. 

Population-based studies are essential to establish a comprehensive and robust understanding of the influence of radio AGNs upon star formation. Previous investigations at low redshifts ($z\lesssim1$) have firmly established that radio AGNs tend to reside in massive, quiescent elliptical galaxies with little ongoing star formation (e.g., \citealt{Best+05}; \citealt{Hickox+09}; \citealt{Heckman+14}). However, at higher redshifts ($z\gtrsim1-1.5$), where galaxies are more likely to host AGNs (e.g., \citealt{Merloni+04}; \citealt{Brandt+15}), there is evidence that radio AGNs can also be found in galaxies with strong star formation. For example, \cite{Kalf+17}, using 74 radio-loud quasars, 72 radio-quiet quasars, and 27 radio galaxies at $0.9 < z < 1.1$, showed that radio-loud quasars have higher star formation rates than radio galaxies. Similarly, \cite{Falkendal+19} studied 25 radio galaxies at $1 < z < 5.2$ and found at least four near the main sequence, while \cite{Gilli+19} discovered a radio galaxy at $z = 1.7$ with strong star formation. These earlier studies, however, were limited by their sensitivity, which restricted them to detecting only the brightest radio AGNs or only radio AGNs within a narrow redshift range.

Recent deep and medium-to-wide radio surveys, such as the 3 GHz VLA COSMOS survey (e.g., \citealt{Smolcic+17}, \citealt{Delvecchio+22}), the Low Frequency Array Two-meter Sky Survey (LoTSS) (e.g., \citealt{Best+23}), and the MIGHTEE survey (e.g., \citealt{Heywood+22}), have enabled the construction of more complete and representative samples of radio AGNs over a broader redshift and luminosity range (e.g., \citealt{Hardcastle+19}; \citealt{Zhu+23}; \citealt{Wang+24}). These new data, combined with abundant multiwavelength observations, provide an opportunity to explore the star-forming characteristics of radio AGN host galaxies in greater detail and over a wider range of cosmic epochs (e.g., \citealt{Igo+24}; \citealt{Wang+24}). To evaluate the preference of AGNs for star-forming or quiescent galaxies, many have used the star-forming fraction ($f_{\text{SF}}$; the fraction of star-forming galaxies to all galaxies) or the incidence ratio (the ratio of AGN fraction in star-forming galaxies to that of quiescent galaxies). Various studies have shown the incidence ratio of \mbox{X-ray} AGNs in star-forming galaxies to be higher than for quiescent galaxies (e.g., \citealt{Aird+19onXrayIncidence}; \citealt{Birchall+22onXrayIncidence}; \citealt{Cristello+24}; \citealt{Zou+24}) out to $z\approx3$. For radio AGNs, which generally represent a distinct population from \mbox{X-ray} AGNs (e.g., \citealt{Hickox+09}), the results remain restricted to a small redshift range, and there has not been a consensus, with cases showing strong preference or no particular preference for radio AGNs to exist in star-forming galaxies. For example, \cite{Best+05} used a sample of 2215 radio AGNs at $0.03<z<0.3$ from the Sloan Digital Sky Survey and found they are more likely to reside in non-star-forming, ``red and dead" galaxies. On the contrary, \cite{Igo+24} identified 682/21462 mass-complete radio AGNs from the GAMA09 galaxies at $z<0.4$ and found that these AGNs do not show a preference for galaxies with low star-formation rates (SFRs). They also found an elevated incidence of AGNs at high $M_{\star}$ and jet-power values. 




Radio AGNs in star-forming and quiescent galaxies may have completely different physical origins (e.g., \citealt{Kondapally+22}). Thus, studying the star-forming and quiescent populations separately is a valuable approach. Star-forming galaxies are known to follow a tight correlation between $M_{\star}$ and SFR, which is referred to as the main sequence (MS; e.g., \citealt{Whitaker+12}; \citealt{Popesso+19}; \citealt{Leja+22}). Thus, for the star-forming population of radio AGNs, we can further probe a more subtle aspect, their distances relative to the MS. 

Over the past decade, there have been numerous efforts to investigate whether the hosts of X-ray AGNs lie above or below the MS at higher redshifts (e.g., \citealt{Mullaney+15}; \citealt{Mountrichas+22, Mountrichas+24}; \citealt{Cristello+24}). \cite{Heckman+14} have shown that most ``jet-mode" AGN hosts in the local universe are quiescent, and for those in transitioning or star-forming galaxies, their host-galaxy SFRs are generally $\approx0.5$ dex lower than the MS value. Local ``radiative-mode" AGNs typically reside in star-forming galaxies but still exhibit similarly lower SFRs than the MS value. Importantly, the populations of X-ray and radio AGNs rarely overlap, highlighting their different properties. Since radio AGNs are mostly ``jet-mode" AGNs, this suggests that radio-AGN host galaxies in the local universe are usually quiescent, contrary to the situation for \mbox{X-ray} AGNs. In the rare cases where radio AGNs are in star-forming galaxies, they have lower SFRs than MS galaxies. Other studies have also firmly established that local radio AGNs have lower SFRs than $M_{\star}$-matched MS galaxies (e.g., \citealt{Gurkan+15}; \citealt{Pace+16}). However, such population analyses are largely missing for radio AGNs at high redshifts, except for some works on individual sources (e.g., \citealt{Markov+20}).



In this work, we leverage radio AGN and galaxy samples from the Wide Chandra Deep Field-South (W-CDF-S), European Large-Area Infrared Space Observatory Survey-S1 (ELAIS-S1), and XMM-Newton Large-Scale Structure (XMM-LSS) fields. These fields benefit from extensive multiwavelength coverage (e.g., \citealt{Zou+22}), which is critical for finding counterparts, determining reliable photometric redshifts and SFRs, and selecting radio AGNs (e.g., \citealt{Zou+21a, Zou+21b, Zou+22}; \citealt{Zhu+23}). Our primary goals are to investigate the fraction of star-forming galaxies among radio AGNs, compare their SFRs to those of MS galaxies, and examine how these properties depend upon $z$ and $M_{\star}$. The sensitivity of our results to the radio AGN selection method is explored. Besides the primary sample from \cite{Zhu+23}, which comprises 1805 rigorously selected radio AGNs detected down to a depth of around $\text{RMS}\approx6-15\:\mu\text{Jy}\:\text{beam}^{-1}$ at 1.4 GHz, for the first time in these fields, we selected additional radio AGN samples with different completeness and purity determined through the radio-infrared correlation.

The layout of this paper is as follows. We describe our multiwavelength data, radio AGN selection, and star-forming galaxy selection in \mbox{Section \ref{sec: data and sample}}. \mbox{Section \ref{sec: analysis}} presents analyses of our sample and relevant discussions, including the fraction of star-forming radio AGNs and the position of radio AGNs relative to the MS. We then examine whether different MS definitions and AGN selections influence our results in this section and Appendix A. The main conclusions are summarized in \mbox{Section \ref{sec: conclusion}}. 

We adopt a flat $\Lambda$CDM cosmology with $H_0 = 70 \:\text{km}\:\text{s}^{-1}\:\text{Mpc}^{-1}$ ,  $\Omega_{\Lambda}=0.7$ and  $\Omega_{\text{M}} = 0.3$ in this paper. The spectral index is defined as $\alpha$ in $f_{\nu}\propto\nu^{\alpha}$. 

\section{Data and Sample} \label{sec: data and sample}

Sources in our study come from three distinct sky fields: W-CDF-S, ELAIS-S1, and XMM-LSS. These fields have rich multiwavelength data and are also the upcoming Vera C. Rubin Observatory Legacy Survey of Space and Time (LSST; \citealt{Ivezic+19}) Deep-Drilling Fields (DDFs; e.g., \citealt{Brandt+18}; \citealt{Zou+22}). These fields also have deep radio coverage. The Australia Telescope Large Area Survey (ATLAS) has obtained radio coverage of the W-CDF-S and ELAIS-S1 fields (\citealt{Norris+06}; \citealt{Hales+14}; \citealt{Franzen+15}), while a VLA survey and the MIGHTEE survey cover the XMM-LSS field (\citealt{Heywood+20, Heywood+22}). We utilize 1.4 GHz observations from these surveys in this study. The ATLAS/W-CDF-S, ATLAS/ELAIS-S1, VLA/XMM-LSS, and MIGHTEE/XMM-LSS surveys cover areas of 3.6, 2.5, 5.0, and $3.5$ $\text{deg}^2$ and reach sensitivities of 14, 17, 16, and 5.6 $  \mu\text{Jy}\: \text{beam}^{-1} $, respectively. The ATLAS/W-CDF-S and ATLAS/ELAIS-S1 surveys have angular resolutions of 16 arcsec $\times$ 7 arcsec and 12 arcsec $\times$ 8 arcsec, respectively. The VLA/XMM-LSS and MIGHTEE/XMM-LSS surveys have angular resolutions of 4.5 and 8.2 arcsec, respectively.
In total, the ATLAS and VLA coverage of the three fields contains $\approx 11000$ radio components, while the deeper MIGHTEE coverage detects more than 20000 radio sources in the XMM-LSS field. \cite{Zhu+23} have compiled 20406 radio sources in their work with data from these fields, from which we utilize the Spitzer MIPS $24\:\mu\text{m}$ flux density ($S_{24\mu\text{m}}$)  and the 1.4 GHz flux density ($S_{1.4\text{GHz}}$) data. Their MIPS $24\:\mu\text{m}$ data originate from the HELP project (\citealt{Shirley+21onHELP}), which deblends confused far-infrared (FIR) sources using the XID+ algorithm. As explained in the Appendix D of \cite{Zhu+23}, the $24\:\mu\text{m}$ fluxes are corrected by factors of 1.11, 0.792, and 1.01 for W-CDF-S, ELAIS-S1, and XMM-LSS, respectively, to bring them in line with other MIPS catalogs: SWIRE (\citealt{Surace+05onSWIRE}), Spitzer Enhanced Imaging Products (SEIP; \citealt{HershelGroup+20}), and Spitzer Data Fusion (\citealt{Vaccari+15onSpitzerData}). In cases where the Bayesian $P$-value residual statistic of the XID+ fit exceeds 0.5, indicating large residuals and unreliable fluxes, \cite{Zhu+23} instead used the MIPS $24\:\mu\text{m}$ fluxes from the Spitzer Data Fusion. We have adopted these fluxes for our analysis as well.
\cite{Zhu+23} also used the DES DR2 catalog in the optical/NIR (\citealt{Abbott+21}), VIDEO DR5 catalog in the NIR (\citealt{Jarvis+13}), and DeepDrill/SERVS IRAC1 catalog in the MIR (\citealt{Lacy+21}) for cross-matching of the radio objects to acquire their precise locations. The VIDEO survey is the best for matching considering its angular resolution and depth, but misses some sources in the W-CDF-S field, while DES and DeepDrill/SERVS provide near-complete coverage (\citealt{Zhu+23}). We refer readers to \cite{Zhu+23} for details of the cross-matching.

The galaxy properties of these radio objects are well characterized and cataloged in \cite{Zou+22} by fitting the corresponding \mbox{X-ray}-to-FIR spectral energy distributions (SEDs) with {\fontfamily{lmtt}\selectfont CIGALE} v2022.0 (\citealt{CIGALEBoquien+20}, \citealt{CIGALEYang+20}, \citealt{CIGALEYang+22}), where AGN contributions were properly considered. We refer readers to \cite{Zou+22} for more details of the SED fitting. The SEDs are generally dominated by galaxy light and suffer little from AGN contamination, which ensures a general high quality of SED fits and photometric redshifts. The redshifts of these radio sources are either spectroscopic or high-quality UV-to-mid-infrared (MIR) photometric redshifts (photo-$z$s) from \cite{Chen+18} and \cite{Zou+21b}. The fraction of spectroscopic redshifts for radio sources in the W-CDF-S, ELAIS-S1, and XMM-LSS fields are 42.3\%, 51.2\%, and 27.5\%, respectively. The photo-$z$ catalogs in \cite{Chen+18} and \cite{Zou+21b} contain a reliability parameter, $Q_z$ (see Section~4.2 of \cite{Brammer+08} for the original definition). Photo-$z$s with $Q_z<1$ is generally reliable, and about 75\% of our photo-$z$s satisfy this. Following \cite{Zou+21b}, we define $\Delta z=z_{\text{phot}}-z_{\text{spec}}$, $\sigma_{\text{NMAD}}$ as the normalized median absolute deviation, and ``outliers" as objects with $|\Delta z|/(1+z_{\text{spec}})>0.15$. For 5594 sources with both spec-$z$s and $Q_z<1$ photo-$z$s, $\sigma_{\text{NMAD}}=0.034$, the outlier fraction is 4.1\%, and the median $\Delta z/(1+z_{\text{spec}})=-0.014$. These values are consistent with those reported for all sources in the W-CDF-S, ELAIS-S1, and XMM-LSS fields (e.g., \citealt{Chen+18, Zou+21b}).



\subsection{Selection of the Main Radio-AGN Sample} \label{subsec: main AGN}

This subsection describes the process of constructing our radio AGN sample. Identifying radio AGNs solely based on their radio morphological structures is challenging. A common approach to overcome this difficulty is to conduct a multiwavelength study by incorporating data from other wavelengths (e.g., \citealt{Magliocchetti+22}). For instance, utilizing FIR data in conjunction with radio data is a suitable method because FIR emission traces star-forming activity. The infrared-radio correlation (IRRC), which has been found to extend over five orders of magnitude with a small dispersion (e.g., \citealt{Yun+01}), has been widely used to distinguish between galaxies powered by a radio-active AGN and star-forming galaxies (e.g., \citealt{Donley+05}; \citealt{Del+13}; \citealt{Bonzini+15}; \citealt{Delvecchio+17}). In this approach, galaxies that deviate significantly from the IRRC in the radio-strong direction are likely to host an AGN component contributing to the radio emission, allowing for the identification of radio AGNs. By combining radio and FIR data, this method provides a reliable means of constructing a radio AGN sample, which is crucial for our analysis.

Radio AGNs, by definition, are selected as sources with characteristic AGN properties in the radio band. \cite{Zhu+23} employed three criteria to select radio AGNs: radio morphology, flat radio spectral slopes, and/or excess radio fluxes. Most of the radio AGNs are selected using the \mbox{radio-excess} method. The first criterion identifies extended jets and lobes, as star-forming galaxies without an AGN typically produce radio emission confined within the galaxy. The second criterion selects sources with radio spectral index $\alpha_r>-0.3$, because star-formation-related radio emission typically has $\alpha_r\approx-0.7\:\mathrm{to}-0.8$ (e.g., \citealt{An+21}). The last criterion utilizes the IRRC: it selects radio AGNs as those with radio flux densities exceeding by tenfold or more those predicted by the $q_{24}$ parameter of \cite{Appleton+04}, where $q_{24}=\log{(S_{24\mu \text{m}}/S_{\text{1.4GHz}})}$.
\cite{Zhu+23} identified 713, 275, and 827 radio AGNs in the W-CDF-S, ELAIS-S1, and XMM-LSS fields, respectively, adding to 1815 radio AGNs in total. The remaining W-CDF-S, ELAIS-S1, and XMM-LSS radio sources are mainly star-forming galaxies and radio-quiet AGNs in W-CDF-S, ELAIS-S1, and XMM-LSS, respectively. 

\cite{Zhu+23} matched the radio-source catalog with the source catalog in \cite{Zou+22}. The \cite{Zou+22} catalog requires VIDEO detection, while the radio source catalog contains sources not detected by VIDEO, so these sources are not cataloged in \cite{Zou+22}. Specifically, 20001/20406 radio objects and 1763/1815 radio AGNs are matched to the catalog of \cite{Zou+22}; thus, we only use the matched sources for analysis. This procedure removes less than 3\% of our total sources and does not have an appreciable impact on our results. We also require the best-fit reduced chi-square ($\chi_r^2$) of the SED-fitting of these AGNs to be less than 5 to remove poor fits. Less than $3\%$ of sources are removed after this procedure, and 1718 are left. Among these radio AGNs, 156 (9.1\%) can be identified by radio morphology, 56 (3.3\%) by radio slope, and 1712 (99.7\%) by radio excess. Due to the small number of morphology-selected radio AGNs, we cannot identify enough Fanaroff-Riley type I (FR I) and type II (FR II) galaxies for comparative study.

These radio AGNs have high-quality measurements of redshifts and host-galaxy properties. The fraction of spectroscopic redshifts for these radio AGNs is 34.1\%, and 78.0\% of the photometric redshifts have quality $Q_z<1$. For 569 sources with both spec-$z$s and $Q_z<1$ photo-$z$s, $\sigma_{\text{NMAD}}=0.025$, the outlier fraction is 3.2\%, and the median $\Delta z/(1+z_{\text{spec}})=-0.013$. The \mbox{FIR-to-UV} emission of radio AGNs is usually galaxy-dominated (\citealt{Zhu+23}), so there is almost no AGN contamination in the optical and infrared bands, ensuring high photo-$z$ quality. Similarly, these radio AGNs generally prefer the galaxy-only model in the SED fitting (\citealt{Zhu+23}), so the effect of AGN contamination on $M_{\star}$ and SFR estimates is usually minor. The median number of good bands (bands with $\mathrm{SNR}>5$) for the SEDs of these AGNs is 13, and 77.2\% of them have more than eight good bands, indicating generally good SED-fitting quality. 



\subsection{New Radio-AGN Selection} \label{subsec: new AGN selection}
In this subsection, we describe the process of creating an alternate sample of radio AGNs. The previous sample employs strict criteria to ensure its purity, which results in reduced completeness. 
We try loosening the criteria and increasing the completeness for our new sample. We rely on the IRRC (e.g., \citealt{Delvecchio+17, Delvecchio+21, Delvecchio+22}) to conduct the new sample selection. This relation is well established for star-forming galaxies with a small dispersion. Radio AGNs are identified as those exhibiting a strong radio excess (i.e., significantly lower IR-to-radio ratios) that cannot be explained by star formation. To select radio AGNs, we define the infrared-to-radio ratio ($q_{\text{IR}}$) of a galaxy as $q_{\text{IR}}=\log(\frac{L_{\text{IR}}\text{[W]}}{3.75\times10^{12}\text{[Hz]}})-\log(L_{\text{1.4GHz}}[\mathrm{W\: Hz^{-1}}])$. Here, $L_{\text{IR}}$ represents the total IR luminosity over rest-frame $8-1000\:\mu\text{m}$, $3.75\times10^{12}\:\text{Hz}$ is the specific frequency of the FIR band, and $L_{1.4\text{GHz}}$ is the spectral luminosity at rest-frame $1.4\:\text{GHz}$. Radio AGNs can be identified by examining the distribution of $q_{\text{IR}}$ among all radio objects if the radio survey is sufficiently deep to detect a large number of MS galaxies; a tail at the low end of this distribution represents radio AGNs. We compute each radio object's $q_{\text{IR}}$ as previously defined. This process entails the calculation of both $L_{1.4\text{GHz}}$ and $L_{\text{IR}}$. We utilize a spectral index of $\alpha_{\text{r}}=-0.7$ to convert $S_{1.4\text{GHz}}$ from \cite{Zhu+23} into the rest-frame $L_{1.4\text{GHz}}$. To derive $L_{\text{IR}}$, we employ the best-fit SED curves provided by \cite{Zou+22} and integrate over the wavelength range of rest-frame $8-1000\:\mu\text{m}$. Of the 20001 radio sources, 70\% have SNRs $>5$ in at least one FIR band (mainly 24 $\mu$m), and 40\% have SNRs $>5$ in at least one of the Herschel bands. The Herschel bands capture the bulk of the FIR emission from a typical galaxy and can significantly enhance the accuracy in estimating a galaxy's \( L_{\text{IR}} \). Note that $L_{\text{IR}}$ can also be inferred from UV-to-optical SEDs because FIR photons originate from the reemission of absorbed UV photons. Given the rich multiwavelength data covering the UV-to-MIR for our sources, we can still reasonably estimate $L_{\text{IR}}$ even for FIR-undetected sources. For example, \mbox{Figure 29} in \cite{Zou+22} shows that excluding the FIR data for FIR-detected sources would only cause little to zero biases of the SED-fitting results. For both galaxies and AGNs, the median offsets of SED fitting results excluding FIR data are $0.002$ dex and $-0.02$ dex for $M_{\star}$ and SFR, respectively (\citealt{Zou+22}). Thus, SFR can be estimated well for sources without FIR data. Since $L_{\text{IR}}$ is an indicator of SFR, $L_{\text{IR}}$ for our sources can also be well-estimated without FIR data. For sources with reliable 24 $\mu$m flux data, we compare $L_{\text{IR}}$ from two SED fits: one with and one without 24 $\mu$m flux to investigate if they systematically differ. These two $L_{\text{IR}}$ agree well: the median offset of $L_{\text{IR}}$ fitted without FIR data is only around $-0.02$ dex; thus, the estimated $L_{\text{IR}}$ for our sources is generally reliable.

\begin{figure}
    \centering
    \includegraphics[width=0.4\textwidth]{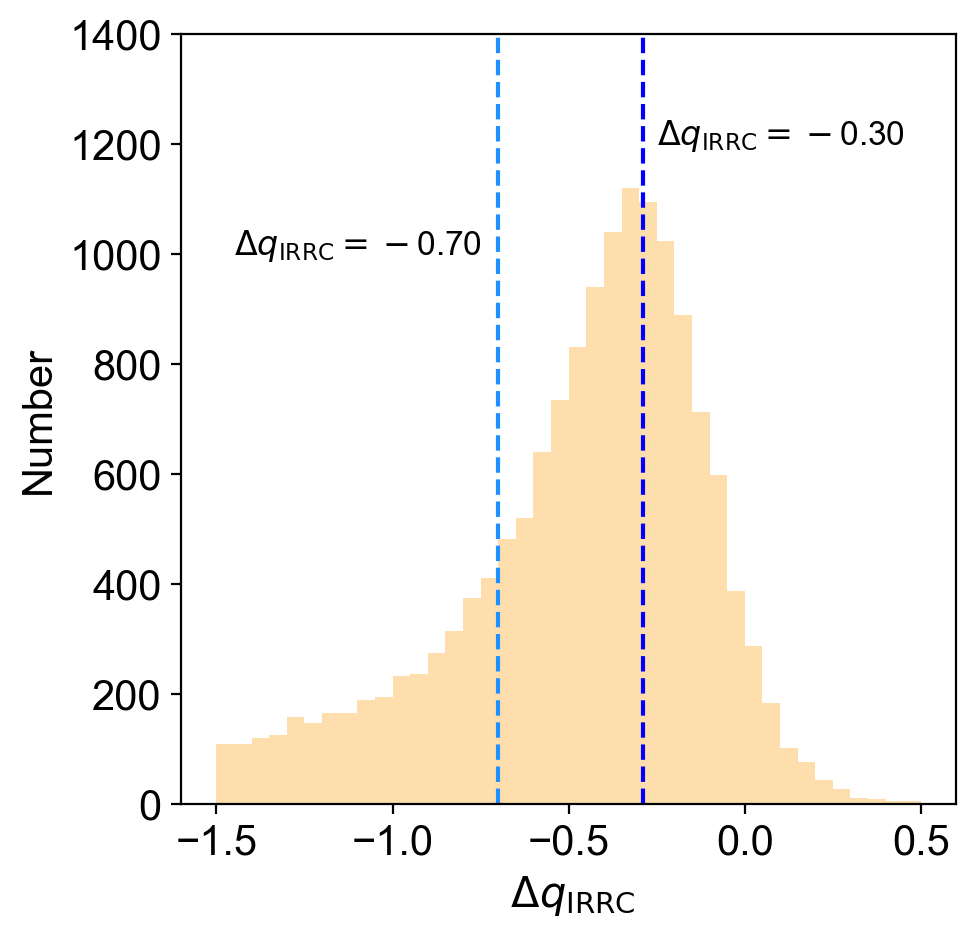}
    \caption{The $\Delta q_{\text{IRRC}}$ distribution of all objects in the XMM-LSS field. The dark blue dashed line shows $\Delta q_{\text{IRRC}}=-0.30$, which corresponds to the peak of the distribution. The light blue dashed line shows the $\text{central}-2\sigma$ cut level ($\Delta q_{\text{IRRC}}=-0.70$).}
    \label{fig 2.2.1}
\end{figure}

\cite{Delvecchio+21} have calibrated the IRRC for star-forming galaxies, which further slightly depends on $M_{\star}$ and $z$. Sources with $q_{\text{IR}}$ much smaller than the IRRC expected value ($q_{\text{IRRC}}$)  are thought to be powered by AGNs in the radio band. We adopt the IRRC in \cite{Delvecchio+21}: $q_{\text{IRRC}}=2.646-0.137\log_{10}{(1+z)}+0.148(\log_{10}{M_{\star}[M_{\odot}]}-10)$ and use this relation to select our sample. Defining $\Delta q_{\text{IRRC}}=q_{\text{IR}}-q_{\text{IRRC}}$, sources with small $\Delta q_{\text{IRRC}}$ can be selected as radio AGNs. We show the distribution of $\Delta q_{\text{IRRC}}$ in \mbox{Figure \ref{fig 2.2.1}}, and its peak is slightly shifted from zero. We argue that such a difference is primarily driven by the prevalent systematic uncertainty of $L_{\text{IR}}$. \cite{Delvecchio+21} and \cite{Zou+22} both measured $L_{\text{IR}}$ through SED fitting, and it is known that SED fitting has an inherent factor-of-two uncertainty (e.g., \citealt{Pacifici+23}). Given that $L_{\text{IR}}$ and SFR scale together (e.g., \citealt{Lutz+14}) and different SED fits often return SFRs systematically differing by $\approx 0.3$ dex, the $\approx 0.3$ dex difference (see below) between the $q_{\text{IRRC}}$ in \cite{Delvecchio+21} and our $q_{\text{IR}}$ is not unexpected. \cite{Delvecchio+21} also started from a mass-complete sample and performed stacking to derive $q_{\text{IRRC}}$, while we start from a radio-selected sample, which naturally have higher radio luminosity and lower $q_{\text{IR}}$. Additionally, the calculations of \cite{Delvecchio+21} are based on the 3 GHz VLA data with a small beam size in the COSMOS field, which might lose a fraction of the radio emission (e.g., \citealt{Zhu+23}). In line with this, \cite{DeZotti+24} suggested that the VLA-COSMOS flux densities might be underestimated by a factor of $\approx 1.51\:(0.18\:\mathrm{dex})$. However, these factors only affect the normalization of the IRRC, meaning our derived $\Delta q_{\text{IRRC}}$ distribution is merely shifted, without impacting the AGN selection. As coefficients in the IRRC relation are the same across fields, we add a correction of $0.3$ dex to the normalization so that the $\Delta q_{\text{IRRC}}$ distribution peaks at 0. 

\begin{figure*}
    \centering
    \includegraphics[width=1.0\textwidth]{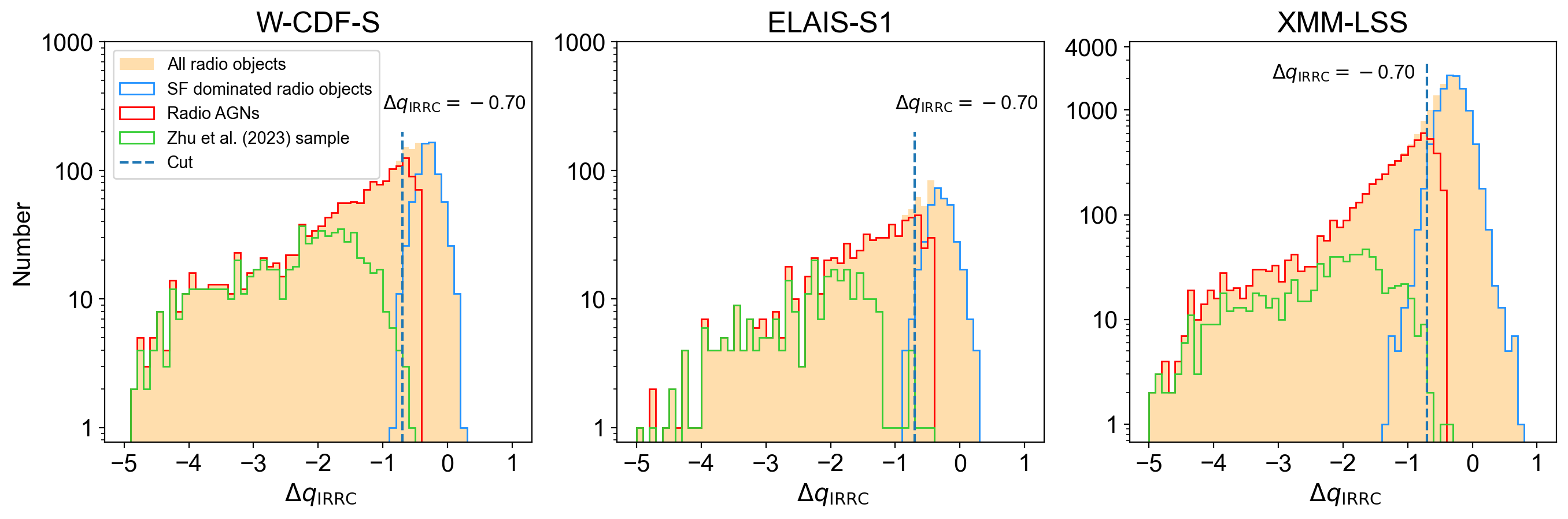}
    \caption{These plots show the $\Delta q_{\text{IRRC}}$ distribution of all objects in each field in orange: W-CDF-S (left), ELAIS-S1 (middle), XMM-LSS (right). We mirror the right part of the distribution to 
the left and assume that it is the distribution of SF-dominated radio objects, which is plotted in blue. We then subtract the two distributions to obtain the radio AGN distribution, which is plotted in red. The blue dashed lines correspond to a $\Delta q_{\text{IRRC}}$ of $-0.70$, which is our selection criterion for radio AGNs. We show radio AGNs selected by \cite{Zhu+23} in green.}
    \label{fig 2.2.2}
\end{figure*}

\begin{figure}
    \centering
    \includegraphics[width=0.4\textwidth]{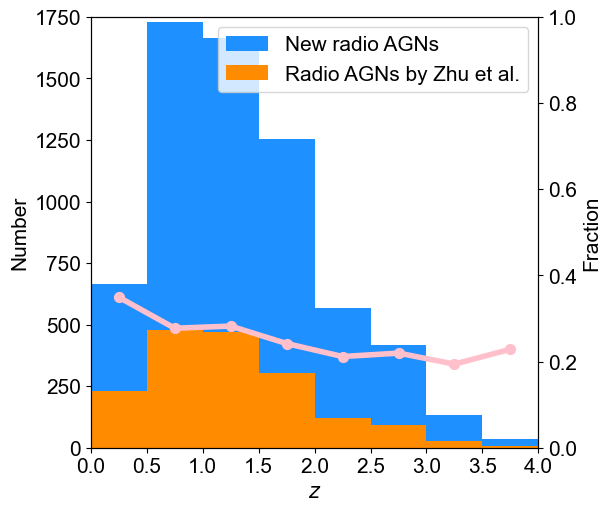}
    \caption{Redshift distributions of the two radio-AGN samples. The redshift distributions of the \cite{Zhu+23} sample and the new sample are displayed in orange and blue, respectively. Additionally, the ratio of the number of sources in the \cite{Zhu+23} sample to that in the new sample is displayed in pink.}
    \label{fig 2.2.3}
\end{figure}

We proceed to analyze the distribution of $\Delta q_{\text{IRRC}}$ within the XMM-LSS field in order to derive the distribution of MS galaxies in the $\Delta q_{\text{IRRC}}$ graph. We perform this analysis only for objects in the XMM-LSS field because the MIGHTEE survey in this field is much deeper than the radio surveys in the other fields, meaning that many more MS galaxies are detected compared to AGNs. This ensures that we obtain the most accurate galaxy distribution data. Following the same procedure as \cite{Delvecchio+22}, who assume that the peak of the $\Delta q_{\text{IRRC}}$ distribution is entirely due to star-forming galaxies and radio-quiet radio AGNs and that the intrinsic distribution of star-forming galaxies is symmetric around the peak (e.g., \citealt{Gurkan+18}), we symmetrize the distribution by reflecting the right half onto the left. We derive the expected histogram representing radio AGNs by subtracting the histogram of MS galaxies from the total one. We then calculate the scatter of MS galaxies as 0.20 dex, which aligns with other works on the IRRC (e.g., \citealt{Yun+01}; \citealt{Delvecchio+21}). To select a large while still reliable radio AGN sample, we aim to identify a threshold that effectively balances sample purity and completeness. We have explored different cut levels, including $\text{central}-1\sigma$, $\text{central}-2\sigma$, $\text{central}-3\sigma$, and $\text{central}-4\sigma$. The outcomes are summarized in \mbox{Table \ref{table 2.2.1}}, in which we estimate the expected purity and completeness using the AGN and MS-galaxy distributions. We adopt the $\text{central}-2\sigma$ criterion, achieving over 95\% purity while missing less than 20\% of the total radio AGNs. The AGN selection results for three fields are shown in \mbox{Figure \ref{fig 2.2.2}}. Employing this selection criterion, we identify 4841 radio AGNs in the XMM-LSS field, utilizing complete FIR and radio data, in contrast to the 797 in the original sample from \cite{Zhu+23}. We apply the same selection criteria to the other fields and identify 1340 and 585 radio AGNs in the W-CDF-S and ELAIS-S1 fields, respectively. This marks a substantial increase compared to the 694 and 272 radio AGNs in the original sample for the respective fields.
In total, we now have 6766 radio AGNs across the three fields, only missing 11 (0.6\%) in the original sample of 1718, which is expected, as we aim to select more AGNs while keeping the original ones. The new sample is nearly three times larger than the previous one. We compare the redshift distribution of two radio-AGN samples in \mbox{Figure \ref{fig 2.2.3}}. These two samples have very similar redshift distributions, both peaking around $z\approx1$. We also present the ratio of the number of sources in the \citet{Zhu+23} sample to that in the new sample for each redshift bin. This ratio is approximately 0.35 at \( z \approx 0 - 0.5 \), decreasing to around 0.2 at \( z \gtrsim 1.5 \). Overall, the ratio remains fairly stable at $z\approx0-4$. A radio-AGN catalog containing the AGNs in \cite{Zhu+23} and in this work is given in \mbox{Table \ref{table 2.2.2}}. The fraction of spectroscopic redshifts for these new radio AGNs is 25.0\%, and 68.9\% of the photo-$z$s have $Q_z<1$. For 1531 sources with both spec-$z$s and $Q_z<1$ photo-$z$s, $\sigma_{\text{NMAD}}=0.031$, the outlier fraction is 6.5\%, and median $\Delta z/(1+z_{\text{spec}})=-0.012$. Although the redshift quality for this new sample is slightly worse than for the main AGN sample described in Section \ref{subsec: main AGN}, it remains satisfactory overall. The quality of the SEDs for these radio AGNs is similarly good to the previous sample, as the median number of good bands is 14, and 76.2\% of them have more than eight good bands.

\begin{table*}
\centering
\caption{\label{table 2.2.1} Expected radio-AGN sample statistics from different criteria.}
\begin{tabular}{ccccccc} \hline\hline
Sample& Total Radio AGNs& W-CDF-S&ELAIS-S1  &XMM-LSS&Expected purity&Expected completeness\\ \hline
 \cite{Zhu+23}& 1763& 694& 272& 797& $>95.6\%$&22.3\%\\ \hline
$\text{Central}-4\sigma$& 3989& 957&  421&2611& 99.7\%&50.2\%\\ \hline
$\text{Central}-3\sigma$& 5072& 1117&  490&3465& 99.1\%&63.4\%\\ \hline 
$\text{Central}-2\sigma$& 6766& 1340&  585&4841& 95.2\%&81.3\%\\ \hline 
$\text{Central}-1\sigma$& 9557& 1639&  700&7218& 80.0\%&96.5\%\\ \hline

\end{tabular}

\end{table*}

\begin{table*}
\caption{Radio AGN selection and science results}
\label{table 2.2.2}
\centering
\begin{threeparttable}
\begin{tabular}{cccccccc} \hline\hline
Field & Name & RA & DEC & Tractor ID & $z$ & $z_{\text{phot, max}}$ & \\
 &  & (degree) & (degree)\\
(1) & (2) & (3) & (4) & (5) & (6) & (7)\\
\hline
W-CDF-S & J032646.45-284952.7 & 51.69349 & $-28.8313$ & 276153 & 0.853 & zphot  \\
W-CDF-S & J032647.93-283142.0 & 51.69967 & $-28.52831$ & 340246 & 0.967 & zphot  \\
W-CDF-S & J032648.14-284329.9 & 51.70053 & $-28.72495$ & 300212 & 0.781 & zphot  \\
W-CDF-S & J032648.20-275747.2 & 51.7008 & $-27.96309$ & 477992 & 1.152 & zphot  \\
W-CDF-S & J032648.36-280003.8 & 51.70147 & $-28.00104$ & 471557 & 1.705 & zphot  \\
\hline
\hline
$z_{\text{phot, min}}$ & $z$-type & \cite{Zhu+23} & New radio AGN & $S_{\mathrm{1.4GHz}}$ & $S_{\mathrm{1.4GHz, \mathrm{err}}}$ & $M_{\star, \mathrm{best}}$ & \\
& & radio AGN & & (mJy) & (mJy) & ($10^9\:M_{\odot}$) & \\
(8) & (9) & (10) & (11) & (12) & (13) & (14)\\
\hline
0.762 & 0.903 & 0 & 1 & 0.334 & 0.041 & 115.713 \\
0.862 & 1.058 & 1 & 1 & 0.632 & 0.04 & 122.726 \\
0.668 & 0.905 & 1 & 1 & 0.298 & 0.034 & 160.813 \\
1.034 & 1.163 & 1 & 1 & 0.895 & 0.05 & 7.645 \\
1.629 & 1.979 & 1 & 1 & 0.18 & 0.021 & 187.871 \\
\hline
\hline

$M_{\star, \mathrm{best, err}}$ & $\mathrm{SFR}_{\mathrm{best}}$ & $\mathrm{SFR}_{\mathrm{best, err}}$ &$N_{\mathrm{good band}}$ & SF\:($UVJ$) & SF\:(nSFR) & $\log L_{\mathrm{IR}}$ & \\
($10^9\:M_{\odot}$) & ($M_{\odot}\:\mathrm{yr}^{-1}$) & ($M_{\odot}\:\mathrm{yr}^{-1}$)&  & & & (W) \\
(15) & (16) & (17) & (18) & (19) & (20) & (21)\\
\hline
37.93 & 11.837 & 11.838 & 15 & 1 & 1 & 38.33078 \\
56.333 & 1.365 & 1.403 & 12 & 0 & 0 & 37.5398 \\
58.093 & 0.29 & 0.43 & 9 & 0 & 0 & 37.29028 \\
2.661 & 60.035 & 21.82 & 10 & 1 & 1 & 38.3326 \\
48.723 & 0.596 & 1.158 & 7 & 0 & 0 & 37.76526 \\

\hline
\hline

$\Delta q_{\mathrm{IRRC}}$ & $\Delta \mathrm{MS}$\:($UVJ$) & $\Delta \mathrm{MS}$\:(nSFR) \\
 \\
(22) & (23) & (24) \\
\hline
$-1.01412$ & 0.16105 & $-0.00781$ \\
$-2.20997$ & $-0.98889$ & $-1.07393$ \\
$-1.938$ & $-1.36519$ & $-1.59538$ \\
$-1.56314$ & 0.78139 & $0.78812$ \\
$-2.02254$ & $-1.99462$ & $-1.9701$ \\

\hline 

\end{tabular}
\begin{tablenotes}
\item
\textit{Notes.} We only show results for the top five rows of our AGN sample here. The full table is available as online supplementary material. Column (1): Field name. Column (2): Object name. Column (3)(4): J2000 R. A. and Decl. Column (5): Tractor ID in \cite{Zou+22}. Column (6): Redshift. Column (7): Redshift type. Column (8)(9): The
68\% lower and upper limit of photo-$z$. These columns are set to $-1$ for sources with spec-$z$s. Column (10): The flag for radio AGNs selected by \cite{Zhu+23}. Column (11): The flag for radio AGNs selected in this paper. Column (12)(13): The flux density and error at 1.4 GHz. Column (14)(15): The best-fit $M_{\star}$ and error from \cite{Zou+22}. Column (16)(17): The best-fit SFR and error from \cite{Zou+22}. Column (18): The number of good bands in the SED fitting. Column (19)(20): The flags for $UVJ$- and nSFR-selected star-forming AGNs. Column (21): $\log L_{\mathrm{IR}}$. Column (22): $\Delta q_{\mathrm{IRRC}}$. Column (23)(24): $\Delta\mathrm{MS}$ for $UVJ$- and nSFR-selected star-forming radio AGNs.
\end{tablenotes}
\end{threeparttable}
\end{table*}

\subsection{Reference-Galaxy Sample} \label{subsec: reference galaxy}
To select our reference-galaxy sample from the cataloged sources in \citet{Zou+22}, we first remove stars from these sources using the {\fontfamily{lmtt}\selectfont "flag\_star"} flag provided in \cite{Zou+22}. We then reject \mbox{X-ray} AGNs and infrared (IR) AGNs from our sample using the {\fontfamily{lmtt}\selectfont "flag\_Xrayagn"}, {\fontfamily{lmtt}\selectfont "flag\_IRagn\_S05"},  {\fontfamily{lmtt}\selectfont "flag\_IRagn\_L07"}, and {\fontfamily{lmtt}\selectfont "flag\_IRagn\_D12"} flags. We do not reject the radio AGNs. For both radio-AGN definitions adopted in this work, these objects constitute less than 0.3\% of our sample and do not have an observable influence on our results.
5.8\% of the objects are rejected, and 2708458 remain after this procedure. Subsequently, we require $\chi_r^2$ of the SED fitting of these sources to be less than 5, following the same criterion as for the radio AGNs. Finally, we have 2664265 objects left. We will utilize these galaxies as a reference sample for investigating the star-forming characteristics of the radio AGNs.

\subsection{Selection of Star-Forming Galaxies and Definition of the MS} \label{subsec: SF selection}

To compare the star-forming characteristics of radio AGNs and galaxies, we need to define and select the star-forming galaxies. Different methods have been used to separate star-forming galaxies from quiescent galaxies, and the dividing line usually becomes dependent upon the adopted selection method when there are too many quiescent galaxies, i.e., at low $z$ and/or high $M_{\star}$. We perform this division in this section and use the star-forming galaxies to establish the MS. The MS has a $0.2-0.3$ dex scatter and holds up to $z\sim6$, with the normalization depending upon $z$ (e.g., \citealt{Speagle+14}, \citealt{Popesso+23}).  The exact shape of the MS depends strongly on the star-forming galaxy definition, and it has been demonstrated that different definitions can produce MS with clearly different shapes, especially at high $M_{\star}$ (e.g., \citealt{Magliocchetti+22}). Thus, it is necessary to test different definitions of star-forming galaxies in this work to ensure the robustness of our results.

\begin{figure*}
    \centering
    \includegraphics[width=1.0\textwidth]{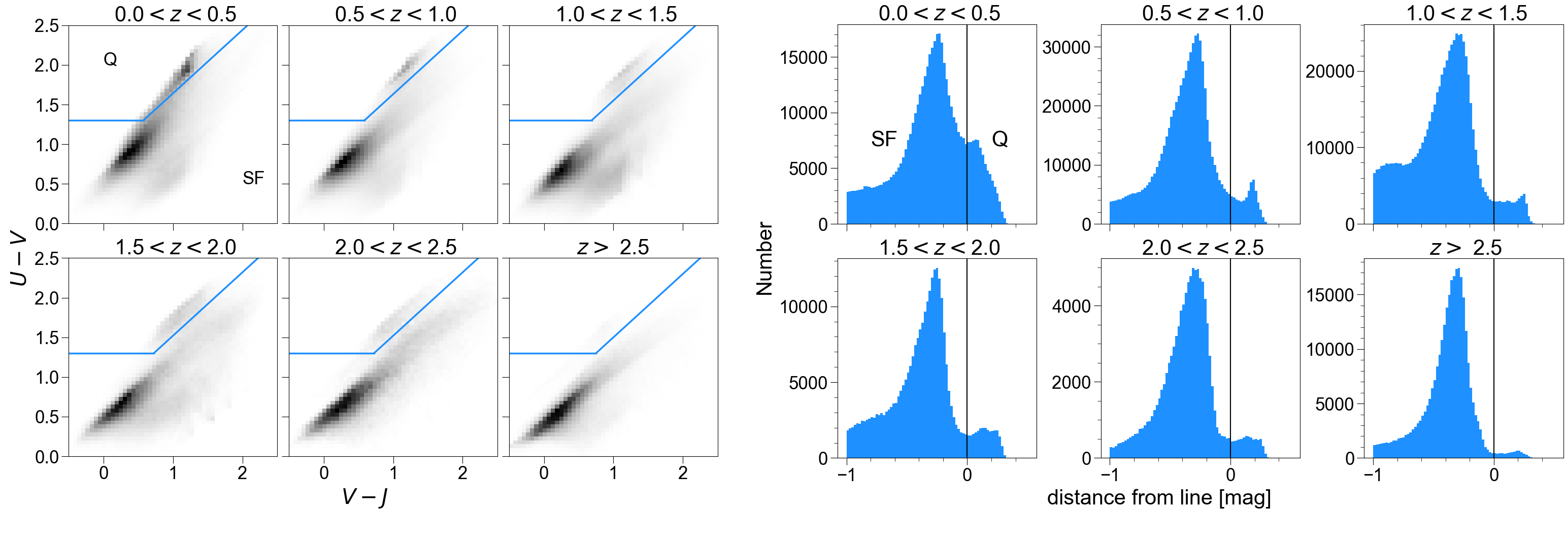}
    \caption{\textit{Left}: Distribution of galaxies in the $UVJ$ color-color space. Darker areas denote regions with a higher galaxy density. The blue lines represent boundaries between the quiescent and star-forming regions (labeled as ``Q" and ``SF" in the figure, respectively) defined in \mbox{Section \ref{subsec: SF selection}}. Quiescent galaxies are identified as those situated to the upper left of the lines, while star-forming galaxies are located outside the boundaries.
    \textit{Right}: number of galaxies as a function of distance to the diagonal separation lines. Points to the lines' upper left and lower right are defined to have positive and negative distances, respectively.}
    \label{fig 2.4.1}
\end{figure*}

We first select star-forming galaxies in the three fields using the rest-frame $UVJ$ diagram (e.g., \citealt{Williams+09}; \citealt{Whitaker+12}; \citealt{Lee+18}). We focus on the $UVJ$-based selection in the main text and will present an alternative selection approach in \mbox{Appendix \ref{appen1}}. Galaxies with blue $U-V$ colors generally have relatively unobscured star formation. However, galaxies showing red $U-V$ colors can be either obscured star-forming galaxies or dust-free quiescent galaxies, and by selecting those galaxies with red $V-J$ colors, obscured star-forming galaxies can be separated from quiescent galaxies, which have blue $V-J$ colors. We show the distribution of all the objects in the $UVJ$ plane in the left panel of \mbox{Figure \ref{fig 2.4.1}}, where bimodality is seen to $z\approx2.5$. Selection of star-forming galaxies using $UVJ$ involves one horizontal cut and one diagonal cut in the $UVJ$ plane (e.g., \citealt{vanDerWel+14}; \citealt{Whitaker+15}). We adopt $U-V<1.3$ from \cite{Whitaker+15} for the horizontal cut. For the diagonal cut, we first use a cut that roughly separates the two populations at all redshifts ($U-V>0.8\times(V-J)+0.7$, the same as \citealt{Whitaker+15}). Then, following the method described in \cite{Williams+09}, we fine-tune the position of the diagonal cut for each $z$ bin, keeping the slope unchanged at 0.8. In the right panel of \mbox{Figure \ref{fig 2.4.1}}, we show the number of galaxies as a function of distance to the diagonal separation lines. Each diagonal cut's position is fine-tuned so that the central line roughly falls between the two peaks. We derive the diagonal cut as

$$U-V>0.8\times(V-J)+ 0.84 \:(0.0<z<0.5)$$
$$U-V>0.8\times(V-J)+0.83 \: (0.5<z<1.0)$$
$$U-V>0.8\times(V-J)+0.75 \: (1.0<z<1.5)$$
$$U-V>0.8\times(V-J)+0.72 \: (1.5<z<2.5)$$
$$U-V>0.8\times(V-J)+0.70 \: (z>2.5).$$


Using our $UVJ$ selection method, 2451637 (92.0\%) of all galaxies are classified as star-forming galaxies. The final separation lines are shown in the left panel of \mbox{Figure \ref{fig 2.4.1}}. We subsequently adopt this subset as our primary sample for studying star-forming galaxies. 

\begin{figure*}
    \centering
    \includegraphics[width=1.0\textwidth]{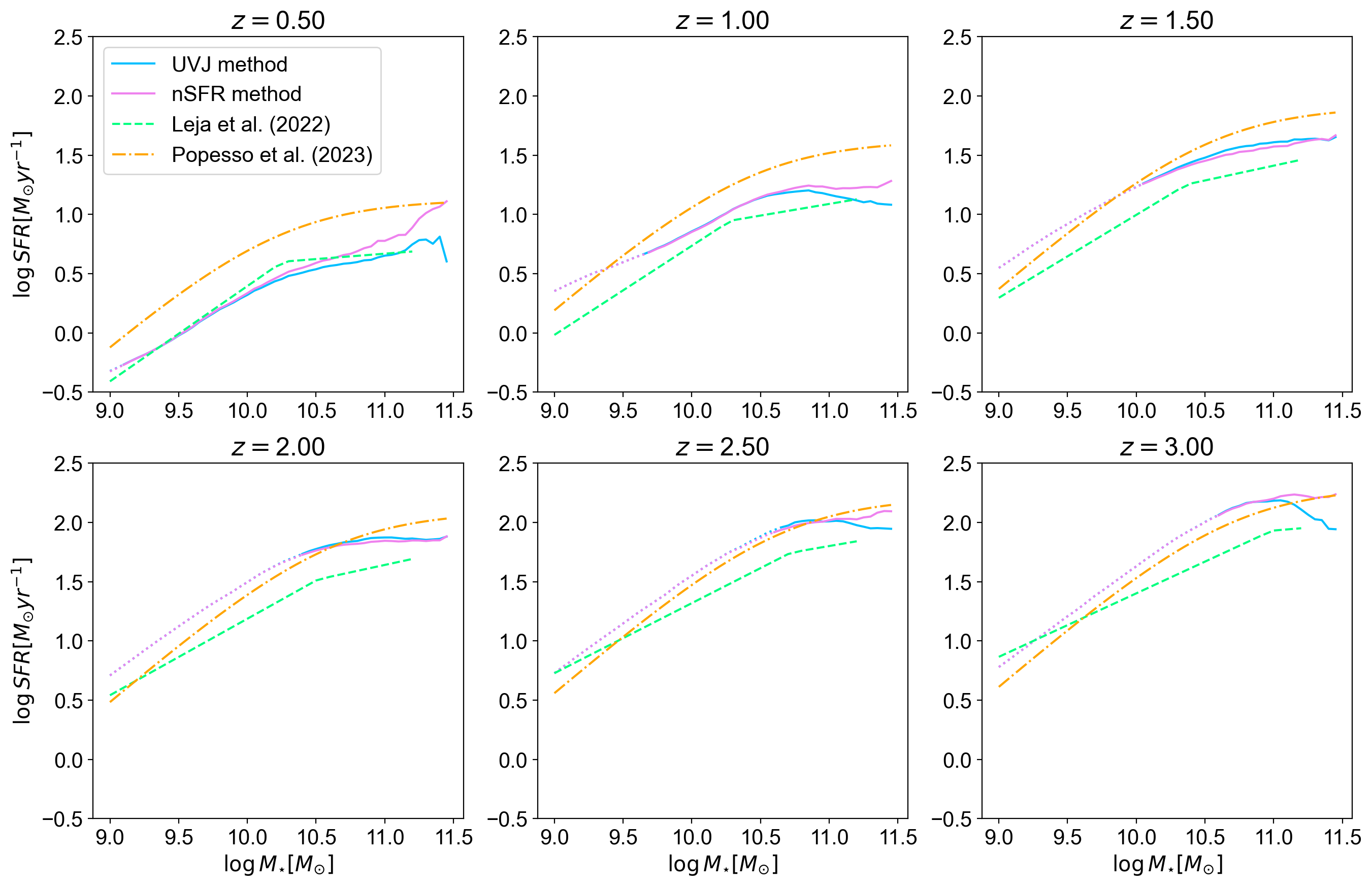}
    \caption{The star-forming MS at different redshifts. The MS determined by $UVJ$- and nSFR-selected star-forming galaxies are shown as blue and violet lines, respectively. The MS in the mass-complete and incomplete domains are shown as solid and dotted lines, respectively. We also compare our MS with \citet[yellow lines]{Leja+22} and \citet[green lines]{Popesso+23} at six redshifts. Our MS generally lie between \cite{Leja+22} and \cite{Popesso+23} from $z=0$ to $z=2$.}
    \label{fig 2.4.2}
\end{figure*}

Now that we have a sample of star-forming galaxies, we define the MS in this work by segregating these star-forming galaxies into bins of $M_{\star}$ ($\pm 0.1$ dex) and $z$ ($\pm 0.075\times(1+z)$). The bin sizes of $\pm 0.1$ dex for $M_{\star}$ and $\pm 0.075\times(1+z)$ for $z$ match the typical uncertainties associated with $M_{\star}$ and $z$, respectively. Increasing the bin sizes results in almost no change in our derived MS, demonstrating that it remains robust at the selected bin sizes. We compute the median SFR of the star-forming galaxies for each bin, plotted as a function of $M_{\star}$ and $z$. \mbox{Figure \ref{fig 2.4.2}} shows the relationship between SFR and $M_{\star}$ at six redshifts ($z=$ 0.5, 1.0, 1.5, 2.0, 2.5, 3.0). In \mbox{Figure \ref{fig 2.4.2}}, we also compare our MS with those presented by \cite{Leja+22} and \cite{Popesso+23} and find that our MS generally align with theirs at $z\approx0.5-3.0$ with an offset $\lesssim0.2$ dex. Note that galaxies selected using the $UVJ$ method may include starburst galaxies, which means not all selected galaxies are MS. This could introduce biases into our MS analysis. However, only 0.1\% of all the star-forming galaxies in our sample have SFR over 1~dex higher than the MS, and thus starburst galaxies only constitute a small fraction of our sample. Our MS SFR is based on the median, which is a robust estimator against outliers. Removing these outliers almost has no impact on our derived MS.

We apply the same methods to the \cite{Zhu+23} radio AGN sample and the $q_{\text{IRRC}}$-selected new sample and select 712 (41.4\%) and 4509 (67.5\%) star-forming ones, respectively. This subset is adopted as the sample for investigating the relative position of star-forming radio AGNs to the MS.
AGN emission may affect optical colors and thus undermine the $UVJ$ color selection. However, most of these radio AGNs do not show AGN signatures at other bands except for the radio (\citealt{Zhu+23}). Thus, we assume no contamination for most of the radio AGNs. Admittedly, a tiny fraction of AGNs may have non-negligible contamination. Thus, we also consider the nSFR method, which selects samples of star-forming galaxies and radio AGNs based solely on $M_{\star}$ , SFR and $z$. Our SED-derived $M_{\star}$ and SFR are robust against contamination because the SEDs have properly accounted for AGN contributions. The nSFR MS is also shown in \mbox{Figure \ref{fig 2.4.2}} and will be discussed in detail in \mbox{Appendix \ref{appen1}}, where we conclude that different definitions of star-forming galaxies cause little difference in our conclusions.

\subsection{Mass-complete Samples} \label{subsec: mass complete}
\begin{figure}
    \centering
    \includegraphics[width=0.4\textwidth]{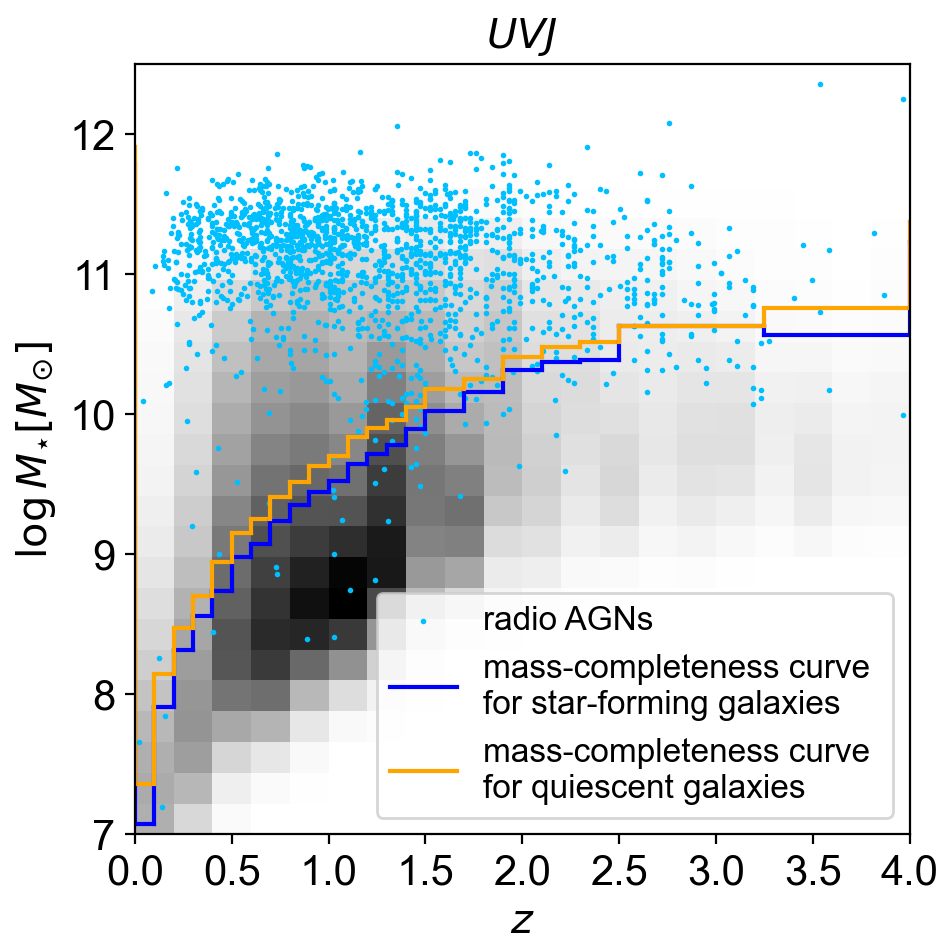}
    \caption{The distribution of radio AGNs and star-forming galaxies in the $\log M_{\star}-z$ space. The blue dots show radio AGNs selected by \cite{Zhu+23}. Darker areas denote regions with a higher galaxy density. The blue and orange lines show the mass-completeness curves for star-forming and quiescent galaxies selected by the $UVJ$ method, respectively. Objects above these lines are defined as mass complete.  } 
    \label{fig 2.3}
    
\end{figure}

We construct complete radio-AGN and reference-galaxy samples by considering the mass-completeness limits. Ensuring mass completeness is crucial for our study, as failing to do so may introduce sample biases. Generally, star-forming galaxies are more easily detected due to their lower mass-to-light ratios. Consequently, the detected quiescent galaxies will be biased toward more massive systems at a given limiting magnitude. In our study, this bias makes comparing the properties of star-forming and quiescent galaxy populations problematic if the sample is not mass-complete. Following \cite{Zou+24}, we define mass completeness using the VIDEO $K_s$-band. We choose the limiting VIDEO $K_s$-band magnitude as $K_{s,\text{lim}}=23.5$, giving a completeness of about 90\% (\citealt{Jarvis+13}).
For each $K_s$-detected object, we calculate its mass limit using the formula: $\log M_{\text{lim}}=\log M_{\star}+0.4(K_s-K_{s,\text{lim}})$ (\citealt{Pozzetti+10}). Then, for each $z$ bin, we define the mass-completeness limit as the value above which 90\% of the mass-limit values lie. We remove sources below this mass-completeness limit, showing our mass-completeness curve in \mbox{Figure \ref{fig 2.3}}. We calculate mass-complete limits for the star-forming and quiescent galaxies separately because the $M_{\star}$ at which quiescent galaxies are considered complete is typically 0.2--0.3 dex higher than for star-forming galaxies. 


When comparing the properties of the star-forming and the quiescent populations, adopting the higher mass limit for quiescent galaxies ensures that both populations are equally complete down to the same $M_{\star}$, allowing an unbiased comparison. If a particular analysis only involves properties of the star-forming population, such as the MS, we can use the lower mass-completeness limit for the star-forming galaxies to keep the sample size as large as possible. Our MS in Section \ref{subsec: SF selection} may be only reliable above this mass-completeness limit, and thus Figure \ref{fig 2.4.2} shows the MS above and below the mass-complete limit as solid and dotted lines, respectively.

If the more stringent mass-completeness limit for quiescent galaxies is used, 869964 (32.7\%) of the galaxies, 1611 (93.8\%) of the \cite{Zhu+23} radio AGNs, and 5862 (88.6\%) of the radio AGNs in \mbox{Section \ref{subsec: new AGN selection}} are classified as mass complete, which shows that most of our radio AGNs lie above the mass-completeness curve. If we use the mass-completeness limit for star-forming and quiescent galaxies separately, the mass-complete ratio is slightly ($<1\%$) higher. Although we lose a fraction of sources in this process, we are still left with a large number of galaxies to define the MS and most of the radio AGNs to study their position relative to the MS. The $M_{\star}$ completeness cut also improves the quality of the redshifts and host-galaxy properties for our sample. Using the \cite{Zhu+23} radio AGNs as an example, the spectroscopic redshift fraction of the mass-complete sample increases to 35.8\%, 1.7\% higher than for the full sample. The fraction of photo-$z$s with $Q_z<1$ rises to 83.6\%, an improvement of 5.6\%. 
Both the galaxy sample and the radio AGN sample in \mbox{Section \ref{subsec: new AGN selection}} exhibit similar enhancements in redshift and measured property quality after applying the $M_{\star}$ completeness criterion. 


\section{Analyses} \label{sec: analysis}

We now have two radio AGN samples and will first concentrate on the primary AGN sample from \cite{Zhu+23} while providing the other results later in this section. We also present the results based on a different selection method of star-forming galaxies in \mbox{Appendix \ref{appen1}} as a comparison. Notably, although the radio depth of the XMM-LSS field is significantly better than for the W-CDF-S and ELAIS-S1 fields, these three fields exhibit nearly identical MS and yield similar outcomes for our analysis. Thus, our results are generally robust under different radio depths. Consequently, we will present the combined result of these fields to enhance the sample size and improve the statistical significance of our findings. Then, we explore results determined by our alternate AGN sample in Section \ref{subsec: effect of AGN}. We also compare the SFR of \mbox{X-ray-} and \mbox{MIR-identified} radio AGNs with those radio AGNs not identified in other bands in \mbox{Section \ref{subsec: HERG}}.

\subsection{Fraction of Star-Forming AGNs} \label{subsec: frac}

The $f_{\text{SF}}$, defined as the fraction of star-forming galaxies to all galaxies, can help us examine the behavior of star-forming characteristics. To ensure a fair comparison of $f_{\text{SF}}$ between star-forming and quiescent galaxies, we adopt the more stringent mass-completeness threshold for the quiescent population, as shown in Figure \ref{fig 2.3} and explained in Section \ref{subsec: mass complete}.

\begin{figure*}
    \centering
    \includegraphics[width=1.0\textwidth]{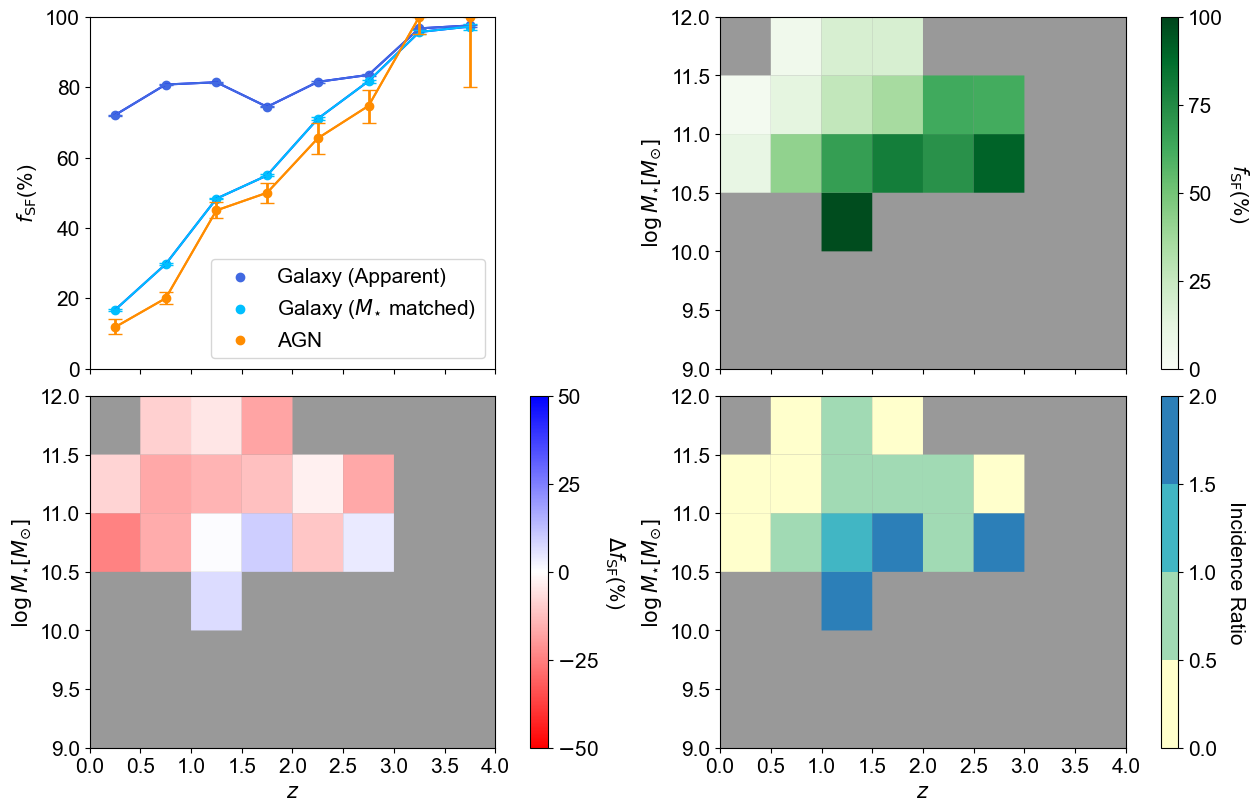}
    \caption{Radio AGNs selected by \cite{Zhu+23} are used in this figure. \textit{Top left}: $f_{\text{SF}}$ of all the galaxies above the mass completeness curves, $M_{\star}$-matched galaxies, and radio AGNs in different bins of $z$, where we plot $f_{\text{SF}}$ of galaxies in royal blue, $M_{\star}$-matched galaxies in deep sky blue, and radio AGNs in orange. The error bars are $1\sigma$-intervals calculated from the binomial proportion confidence intervals. At all redshifts, radio AGNs are more quiescent than galaxies. The divergence is smaller when $z$ increases from 0 to 2. \textit{Top right}: $f_{\text{SF}}$ of AGNs in bins of $z$ and $M_{\star}$. Deeper color represents a higher $f_{\text{SF}}$. The grey region signifies insufficient sample statistics. \textit{Bottom left}: $\Delta f_{\text{SF}}$ of AGNs in bins of $z$ and $M_{\star}$. The blue and red areas indicate higher and lower $f_{\text{SF}}$ than galaxies, respectively. \textit{Bottom right}: Ratio of the measured incidence of radio AGN in star-forming to quiescent galaxies in bins of $z$ and $M_{\star}$. Results are color-coded, and deeper color shows a higher incidence ratio.}
    \label{fig 3.1}
    
\end{figure*}

Adopting this mass-completeness limit, we proceed to calculate $f_{\text{SF}}$ for each $z$ bin for radio AGNs and galaxies in our sample. We show the results in the top-left panel of \mbox{Figure \ref{fig 3.1}}.  The error bars are computed from the binomial proportion confidence intervals. At $z\approx0-0.5$, 76\% of galaxies but only 11\% of radio AGNs are star-forming. The divergence ($\Delta f_{\text{SF}}$, defined as $f_{\text{SF, AGN}}-f_{\text{SF, galaxy}}$) is smaller at higher redshifts, as $82\%$ of galaxies and $64\%$ of radio AGNs are star-forming at $z\approx2-2.5$. $f_{\text{SF}}$ above $z=3.0$ shows large scatters due to a limited sample size but appears to be nearly unity. Since radio AGNs tend to reside in massive galaxies (e.g., \citealt{Best+05}), and $f_{\text{SF}}$ of galaxies declines rapidly with $M_{\star}$, the low $f_{\text{SF}}$ of radio AGNs may be mainly driven by either their high $M_{\star}$ or the presence of radio AGNs. To probe this, we match galaxies within $ 0.1$ dex of $M_{\star}$ and within $ 0.075\times(1+z)$ of $z$ for each AGN. A number of 100 galaxies is randomly selected with replacement from the matched galaxies for each AGN. Then, we combine the selected galaxies for each AGN and plot their $f_{\text{SF}}$ with $z$ also in the top-left panel of \mbox{Figure \ref{fig 3.1}}. This procedure allows us to directly probe the effect of the presence of radio AGNs on $\Delta f_{\text{SF}}$. It appears that although the apparent galaxy $f_{\text{SF}}$ is much higher than for radio AGN at $z\approx0-2$, the $M_{\star}$ matched-galaxy $f_{\text{SF}}$ is only moderately larger than the radio-AGN $f_{\text{SF}}$ by $\lesssim10\%$, indicating that the low radio-AGN $f_{\text{SF}}$ is primarily caused by their high $M_{\star}$, while their AGN nature only plays a secondary role.
To summarize, radio AGNs are far more quiescent than galaxies at low redshifts and mostly reside in quiescent systems at $z\lesssim0.5$; however, the radio-AGN $f_{\text{SF}}$ quickly increases with $z$ and gradually catches up to the galaxy $f_{\text{SF}}$ at $z\lesssim3$. The primary reason for the initially low $f_{\text{SF}}$ of radio AGNs is their tendency to reside in massive galaxies, which inherently have significantly lower $f_{\text{SF}}$ compared to their low-mass counterparts. After accounting for the $M_{\star}$ of radio AGN hosts and matching it to galaxies, the $f_{\text{SF}}$ of radio AGNs becomes nearly comparable to that of galaxies. \cite{Heckman+14} have shown that most ``jet-mode" AGNs in the local universe are in quiescent galaxies (see their Figure~2); these radio AGNs typically have high $M_{\star}$, and at similar $M_{\star}$, most galaxies are also quiescent. Our results align with this picture in the local universe. At higher redshifts, as radio AGNs become less preferential toward residing in high-$M_{\star}$ galaxies, their $f_{\text{SF}}$ increases to a value close to that of galaxies.

We then look further at the dependence for $f_{\text{SF}}$ of radio AGNs on both $z$ and $M_{\star}$. As illustrated in the top-right panel of \mbox{Figure \ref{fig 3.1}}, $f_{\text{SF}}$ exhibits an increasing trend with $z$ and a decreasing trend with $M_{\star}$ for these AGNs. To ensure statistical robustness, we impose a constraint that the 3$\sigma$ uncertainty on $f_{\text{SF}}$ within a given bin must be smaller than 0.25 for the bin to be displayed in the figure. Star formation in high-redshift and massive galaxies containing radio AGNs has been quenched, while low-redshift and smaller galaxies with radio AGNs still tend to be star-forming. This result coincides with what we usually find for galaxies (e.g., \citealt{Martis+16onSFfraction}). To compare further the $f_{\text{SF}}$ of radio AGNs with galaxies in $z$ and $M_{\star}$ bins, we show $\Delta f_{\text{SF}}$ in bins of $z$ and $M_{\star}$ in the bottom-left panel of \mbox{Figure \ref{fig 3.1}}. In most bins, radio AGNs have slightly lower $f_{\text{SF}}$ than galaxies. However, radio AGNs show similar or elevated $f_{\text{SF}}$ in some high $z$ and low $M_{{\star}}$ bins. We further demonstrate this in the bottom-right panel, where the ratio of measured incidence of radio AGN in star-forming to quiescent galaxies ($P_{\text{SF}}/P_{\text{Q}}$, $P_{\text{SF}}$ and $P_{\text{Q}}$ are the incidence of radio AGNs in star-forming and quiescent galaxies) is shown. Many bins show an incidence ratio of less than one but higher than 0.5, and two bins shows an incidence ratio higher than 1. Radio AGNs generally are more likely to reside in quiescent galaxies, but in some parameter regions, they may prefer to live in star-forming galaxies. 

\subsection{Position of Radio AGNs Relative to the MS \label{subsec: deltams}}

From \mbox{Section \ref{subsec: frac}}, we know a considerable fraction of high-redshift radio AGNs are star-forming. To study the position of these star-forming radio AGNs relative to the MS, for each star-forming radio AGN we select star-forming galaxies within 0.1 dex of $M_{\star}$ and within $0.075\times(1+z)$ of $z$. The median number of galaxies matched to each AGN is $\approx5100$. We exclude those AGNs with lower than 100 matches and obtain a sample of 591 (97.8\%) well-matched AGNs. We then investigate the SFRs of these AGNs. 

\begin{figure*}
    \centering
    \includegraphics[width=1.0\textwidth]{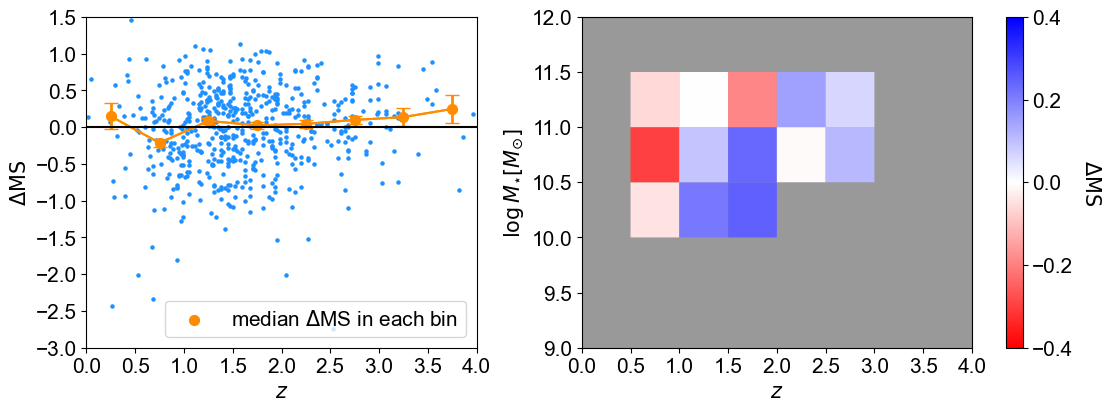}
    \caption{Radio AGNs selected by \cite{Zhu+23} are used in this figure. \textit{Left}: The relationship between $\Delta$MS and $z$ for radio AGNs. Most of the data points are within $0.5<z<2.3$. We have chosen a $z$ bin of 0.4 and plotted the median $\Delta$MS in each bin using orange dots. The error bars are $1\sigma$ uncertainties of the medians. \textit{Right}: Color-coded median $\Delta$MS in different bins of $z$ and $M_{\star}$ for the AGN populations in each bin. The red and blue squares represent bins where radio AGNs are generally below and above the MS, respectively. The grey region signifies insufficient sample statistics.}
    \label{fig 3.2}
\end{figure*}

We calculate these galaxies' median SFR (i.e., $\text{SFR}_{\text{MS}}$) and divide the AGN's SFR by this median SFR to obtain $\Delta \text{MS}$, defined as $\log{(\text{SFR}_{\text{AGN}}/\text{SFR}_{\text{MS}})}$. We plot the radio AGNs' $\Delta \text{MS}$ evolution with $z$ in the left panel of \mbox{Figure \ref{fig 3.2}} and also plot the median values in $z$ bins. We calculated the Pearson correlation coefficient between the variables $\Delta \text{MS}$ and $z$ to be 0.126, which indicates a weak positive linear relationship between the two variables. The associated no-correlation $p$-value of 0.002 suggests the correlation is statistically significant. Additionally, we computed Kendall's tau, a non-parametric measure of the strength and direction of the association between two ranked variables. It has a value of 0.085 with a no-correlation $p$-value $=0.02$, also indicating a weak positive monotonic relationship between $\Delta \text{MS}$ and $z$. The global median $\Delta\text{MS}=-0.03\pm0.03$, consistent with 0 within the uncertainty range. Therefore, star-forming radio AGNs generally lie on or around the MS. \footnote{Given the shape of the stellar mass function, more star-forming galaxies may fall within the range $[M_{\star,\text{AGN}} - 0.1 \:\mathrm{dex}, M_{\star,\text{AGN}}]$ than $[M_{\star,\text{AGN}}, M_{\star,\text{AGN}} +   0.1 \:\mathrm{dex}]$, potentially biasing the sample toward lower stellar masses. At high redshift, galaxies slightly below the AGN's redshift may similarly also be preferentially selected. Since MS SFR increases with both stellar mass and redshift, this could potentially explain the uptick in $\Delta\mathrm{MS}$ with redshift. The median ratio of matched galaxy stellar mass to AGN mass is 0.977, slightly below 1. However, testing narrower bin widths for $M_{\star}$ (0.04 dex and 0.02 dex) showed no impact on our conclusions. The Pearson coefficients for \(\Delta\)MS vs. redshift were 0.12 for both bin widths, the correlations were both significant, and other results also remained unchanged. Similarly, testing narrower redshift bins did not affect our conclusions.}

We then conduct a more in-depth study of the $\Delta \text{MS}$ dependence on $z$ and $M_{\star}$. Color-coded results of the evolution of $\Delta\text{MS}$ with $z$ and $M_{\star}$ are shown in the right panel of \mbox{Figure \ref{fig 3.2}}. We require the number of AGNs in each bin to be greater than 10 to derive reliable results. From \mbox{Section \ref{subsec: frac}}, we know $f_{\text{SF}}$ of radio AGNs is smaller than that of galaxies in most $z$ and $M_{\star}$ bins. However, star-forming radio AGNs have higher SFR than galaxies in over half of the bins. For the distribution of $\Delta \text{MS}>0$ and $\Delta \text{MS}<0$ areas, star-forming radio AGNs with lower SFRs than MS galaxies ($\Delta \text{MS}<0$) primarily reside in the top-left part of the graph, indicating that they are in massive low-$z$ galaxies. Star-forming radio AGNs on or above the MS ($\Delta \text{MS}\geq 0$) can be found both in the lower and right parts of the graph, i.e., in low-mass or high-redshift galaxies. Also seen from the plot is that the $\Delta \text{MS}$ of radio AGNs with $\log{M_{\star}}>11.0$ is generally lower than those with $10.0<\log{M_{\star}}<11.0$, indicating more significant star-formation quenching. 

Overall, a large population of radio AGNs are star-forming at $z\gtrsim1$ and may even lie above the MS at higher redshift, indicating that radio AGNs evolve in parallel to or even exceed the MS galaxy star-formation at high redshift.

\subsection{Comparison of high-$L_{1.4\text{GHz}}$ and low-$L_{1.4\text{GHz}}$ AGNs} \label{subsec: different Lradio}

From \mbox{Section \ref{subsec: frac}}, we already know that the tendency for radio AGNs to reside in more massive galaxies plays a major role in their low $f_{\text{SF}}$. In this subsection, we explore whether the $L_{1.4\mathrm{GHz}}$ of radio AGNs influences their $f_{\text{SF}}$ or $\Delta\text{MS}$. 

When analyzing the relationship between SFR and $L_{1.4\mathrm{GHz}}$ for radio AGNs, a selection effect must be considered. This selection effect arises from the correlation between the SFR and the $24\:\mu\text{m}$ flux ($S_{24\mu\text{m}}$). Because the radio-AGN selection method relies on the ratio of $24\:\mu\text{m}$ flux to $1.4\:\text{GHz}$ flux ($S_{24\mu\text{m}}/S_{1.4\text{GHz}}$), the higher the SFR of a source, the larger $S_{1.4\text{GHz}}$ is required for it to be selected as a radio AGN. However, this bias only affects sources with relatively low radio luminosities, while those radio AGNs can be unquestionably selected if they are sufficiently bright in the radio band such that their strong radio emission cannot be explained by any reasonable SFRs. Therefore, we only analyze these bright radio AGNs to avoid the aforementioned bias. To apply appropriate radio-luminosity cuts, we rely on the MS in \mbox{Section \ref{subsec: SF selection}}, given the fact that galaxies generally can hardly reach $>1$~dex above the MS. For a given $(M_\star, z)$ set, we obtain the MS SFR and derive the rest-frame $S_{24\mu\text{m}}$ that corresponds to ten times the MS SFR (see the next paragraph). In the $q_{24}$ selection process, this $S_{24\mu\text{m}}$ has a corresponding $S_{1.4\text{GHz}}$ threshold for the radio-AGN selection. We thus utilize only sources above this $(M_\star, z)$-dependent $S_{1.4\text{GHz}}$ threshold.

Following \mbox{Section 4.4} of \cite{Kirkparick+12} and assuming a Salpeter IMF (\citealt{Salpeter+55}) and continuous star formation (\citealt{Kennicutt+98}), we have a relation between the obscured component of the SFR ($\text{SFR}_{\text{IR}}$) and the portion of $L_{\text{IR}}$ that is caused by star-formation activity ($L_{\text{IR}}^{\text{SF}}$) as: 

$$\left(\frac{\text{SFR}_{\text{IR}}}{M_{\odot}\:\mathrm{yr}^{-1}}\right)=1.72\times10^{-10}\left(\frac{L_{\text{IR}}^{\text{SF}}}{L_{\odot}}\right).$$

The obscured component constitutes the bulk of star formation in our mass range of interest (e.g., \citealt{Whitaker+17}). Thus, we can link the SFR of a galaxy to its $L_{\text{IR}}^{\text{SF}}$. Typical  star-forming galaxies do not have SFRs beyond ten times of the MS SFR ($\text{SFR}_{\text{MS}}$). Thus, we calculate the limiting $L_{\text{IR}}^{\text{SF}}$ for star-forming galaxies ($L_{\text{IR, lim}}^{\text{SF}}$) as:

$$\left(\frac{10\times\text{SFR}_{\text{MS}}}{M_{\odot}\:\mathrm{yr}^{-1}}\right)=1.72\times10^{-10}\left(\frac{L_{\text{IR, lim}}^{\text{SF}}}{L_{\odot}}\right).$$

Then, we convert $L_{\text{IR, lim}}^{\text{SF}}$ to observed-frame $S_{24\mu\text{m}}$ limits from the \cite{Kirkparick+12} templates. We adopt their $z\approx1$ and $z\approx2$ star-forming-galaxy SEDs for sources at $z<1.5$ and $z\geq1.5$, respectively. Subsequently, we put the $S_{24\mu\text{m}}$ limits back into the $q_{24}$-selection criteria described by \cite{Zhu+23} and obtain the $(M_\star, z)$-dependent $L_{1.4\text{GHz}}$ limit for our sources. We have 869/1613 of our radio AGNs and 416/591 of the star-forming radio AGNs above this limit.

\begin{figure*}
    \centering
    \includegraphics[width=1.0\textwidth]{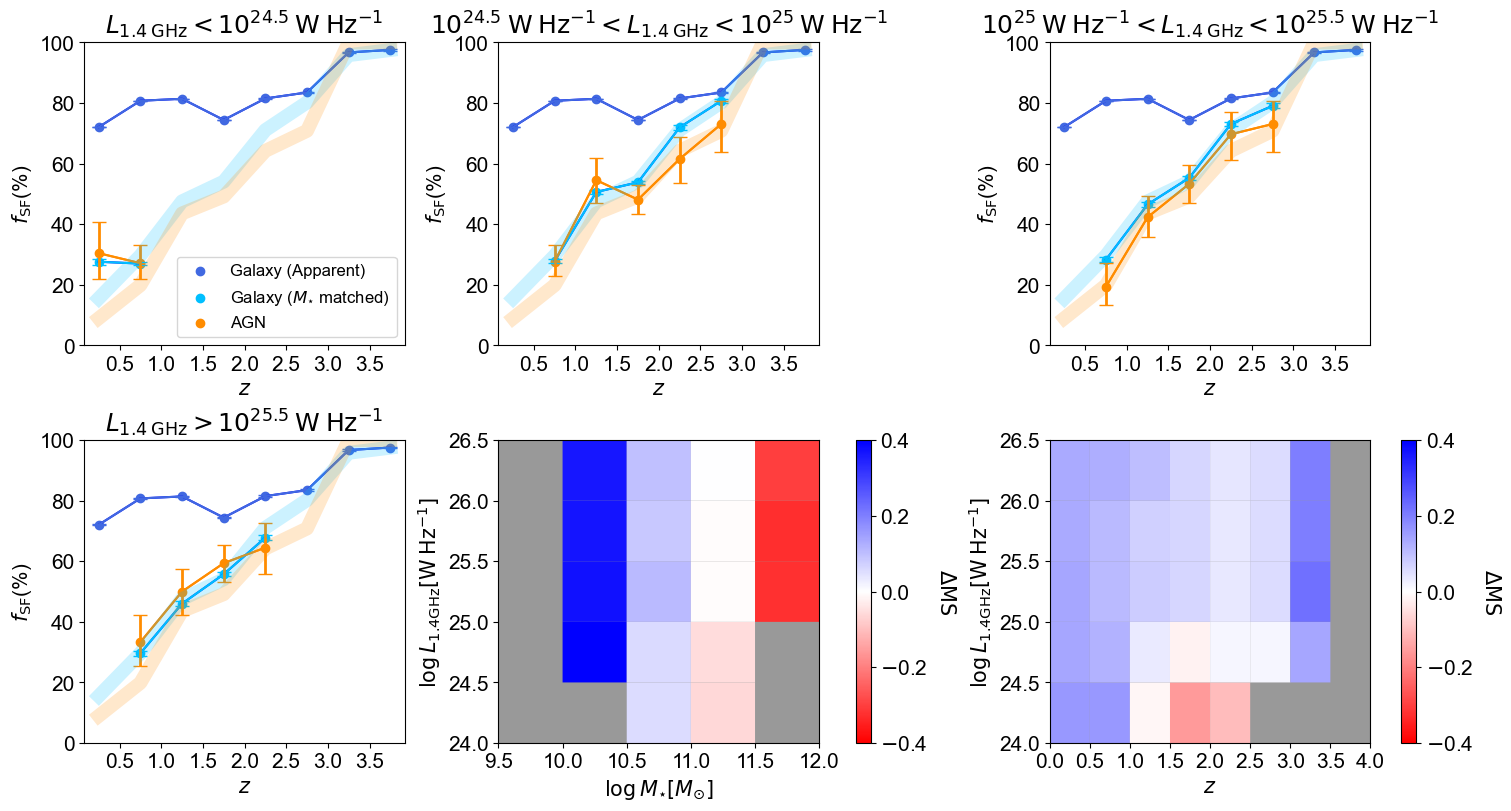}
    \caption{The top-left, top-middle, top-right, and bottom-left panels show the $f_{\text{SF}}$ of galaxies, $M_{\star}$-matched galaxies, and radio AGNs in different $L_\mathrm{1.4GHz}$ bins. Radio AGNs selected by \cite{Zhu+23} are used in this figure. The format of these panels is similar to that of the top-left panel of \mbox{Figure \ref{fig 3.1}}. For comparison, we also added blue and orange transparent lines showing the trends for radio AGNs and $M_{\star}$-matched galaxies from the upper left panel of \mbox{Figure \ref{fig 3.1}}. The bottom-middle panel shows the $\Delta f_{\text{SF}}$ of AGNs in bins of $L_{1.4\text{GHz}}$ and $M_{\star}$. The blue and red areas indicate higher and lower $f_{\text{SF}}$ than galaxies, respectively. The grey region signifies insufficient sample statistics. The bottom-right panel shows the $\Delta\text{MS}$ of AGNs in bins of $L_{1.4\text{GHz}}$ and $z$.}
    \label{fig 3.3.1}
    
\end{figure*}

We first divide our radio-AGN sample into sub-samples with different radio emission levels and calculate their $f_{\text{SF}}$ separately. Most of these AGNs have a $L_{1.4\text{GHz}}$ of $10^{24}\:\text{W}\:\text{Hz}^{-1}-10^{26}\:\text{W}\:\text{Hz}^{-1}$, with a few below $10^{24}\:\text{W}\:\text{Hz}^{-1}$ and $\approx70$ above $10^{26}\:\text{W}\:\text{Hz}^{-1}$. We divide our radio AGNs into four sub-samples: $L_{1.4\text{GHz}}<10^{24.5}\:\text{W}\:\text{Hz}^{-1}$, $10^{24.5}\:\text{W}\:\text{Hz}^{-1}<L_{1.4\text{GHz}}<10^{25}\:\text{W}\:\text{Hz}^{-1}$, $10^{25}\:\text{W}\:\text{Hz}^{-1}<L_{1.4\text{GHz}}<10^{25.5}\:\text{W}\:\text{Hz}^{-1}$, and $L_{1.4\text{GHz}}>10^{25.5}\:\text{W}\:\text{Hz}^{-1}$.  These sub-samples contain 152, 301, 215, and 201 radio AGNs, respectively. We repeat the $f_{\text{SF}}$ calculations in \mbox{Section \ref{subsec: frac}} and show the results in \mbox{Figure \ref{fig 3.3.1}}. The $L_{1.4\text{GHz}}<10^{24.5}\:\text{W}\:\text{Hz}^{-1}$ sub-sample only contains a sufficient number of AGNs at $z<1.0$, and reliable conclusions can only be made in this redshift range.
These radio AGNs generally show similar $f_{\text{SF}}$ compared to $M_{\star}$-matched galaxies, indicating no observable suppression of star formation. Notably, $f_{\text{SF}}$ of radio AGNs remains close to that of galaxies regardless of the value of their $L_{1.4\text{GHz}}$. The $10^{24.5}\:\text{W}\:\text{Hz}^{-1}<L_{1.4\text{GHz}}<10^{25}\:\text{W}\:\text{Hz}^{-1}$ radio AGNs, however, show a little lower $f_{\text{SF}}$ than galaxies at $z>1.5$, indicating a mild suppression of star-formation.  The $10^{25}\:\text{W}\:\text{Hz}^{-1}<L_{1.4\text{GHz}}<10^{25.5}\:\text{W}\:\text{Hz}^{-1}$ radio AGNs generally show similar or slightly lower $f_{\text{SF}}$ when compared to galaxies. Additionally, the $f_{\text{SF}}$ of $L_{1.4\text{GHz}}>10^{25.5}\:\text{W}\:\text{Hz}^{-1}$ radio AGNs is $10\%-20\%$ lower than galaxies at the high-$z$ end. Generally, the difference remains small throughout the $z$ and $L_{1.4\text{GHz}}$ range. Our results show that after considering the tendency for radio AGNs to reside in massive galaxies,  $L_{1.4\text{GHz}}$ does not have a noticeable influence on star-forming activity at $10^{24.5}\:\text{W}\:\text{Hz}^{-1}<L_{1.4\text{GHz}}<10^{25.5}\:\text{W}\:\text{Hz}^{-1}$. 

\begin{figure}
    \centering
    \includegraphics[width=0.4\textwidth]{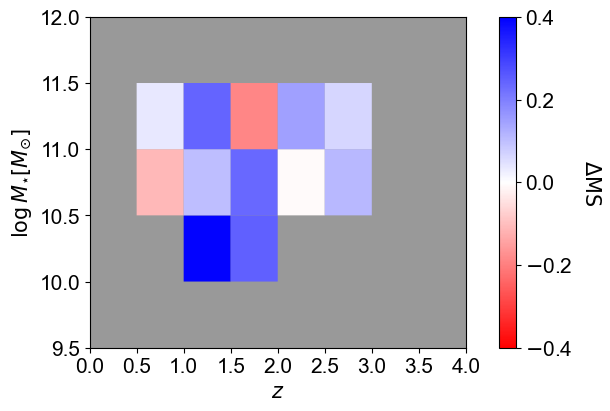}
    \caption{This plot shows the $\Delta\text{MS}$ of radio AGNs brighter than the limiting $L_{1.4\text{GHz}}$ in bins of $L_{1.4\text{GHz}}$ and $z$. These radio AGNs are selected by \cite{Zhu+23}, mass-complete, and above the $(M_\star, z)$-dependent $S_{1.4\text{GHz}}$ threshold. The grey region signifies insufficient sample statistics. }
    \label{fig 3.3.2}
    
\end{figure}

We also calculated $\Delta\text{MS}$ for these radio AGNs in bins of $M_\star$ and $L_{1.4\text{GHz}}$, which is shown in the bottom-middle panel of \mbox{Figure \ref{fig 3.3.1}}. The results are similar to the $f_{\text{SF}}$ findings. The $\Delta\text{MS}$ evolves strongly with $M_{\star}$, but almost does not change with $L_{1.4\text{GHz}}$ at a fixed $M_\star$. The bottom-right panel of \mbox{Figure \ref{fig 3.3.1}} shows $\Delta\text{MS}$ for these radio AGNs in bins of $z$ and $L_{1.4\text{GHz}}$, and $L_{1.4\text{GHz}}$ does not have a significant correlation with $\Delta\text{MS}$ at a fixed $z$. As a reference, \mbox{Figure \ref{fig 3.3.2}} shows the $\Delta\text{MS}$ for these radio AGNs in bins of $M_{\star}$ and $z$. For these radio AGNs that are selected by \cite{Zhu+23}, mass-complete, and above the $(M_\star, z)$-dependent $S_{1.4\text{GHz}}$ threshold, $M_{\star}$ has a weak correlation with $\Delta\text{MS}$ at a fixed $z$. However, the scatter is relatively large. Overall, $L_{1.4\text{GHz}}$ does not have a clear influence on $\Delta\text{MS}$. The instantaneous AGN luminosity might not directly impact the overall SFR of galaxies. This may be explained by the fact that the timescale of AGN feedback required to regulate the star formation is likely longer than that of a single AGN episode (e.g., \citealt{Harrison+21}). 

\subsection{Effect of Different AGN Selections}\label{subsec: effect of AGN}

Using the new radio AGN sample from \mbox{Section \ref{subsec: new AGN selection}}, we performed a similar analysis of the position of radio AGNs relative to the MS. 
The new AGN sample results are shown in \mbox{Figure \ref{fig 4.2.1}}. As discussed in detail in \mbox{Appendix \ref{appen1}}, different MS definitions do not change our results qualitatively, and we will focus on the $UVJ$-defined MS. 

In the top-left and top-right panels of \mbox{Figure \ref{fig 4.2.1}}, we see $f_{\text{SF}}$ of these radio AGNs are generally $5\%-20\%$ higher than for the previous sample, so $\Delta f_{\text{SF}}$ are higher by the same value. The difference is caused by the fact that our new radio-AGN selection in \mbox{Section \ref{subsec: new AGN selection}} is much looser than for \cite{Zhu+23}. These two kinds of selections are both based on the correlation between the radio emission and IR emission for star-forming galaxies, and the strict selection in \cite{Zhu+23} reaches a $>95\%$ purity with the expense of missing many real radio AGNs, while our larger new sample has a higher completeness and more contamination from star-forming galaxies (\mbox{Table \ref{table 2.2.1}}). Nevertheless, although our new sample is three times larger than the original one in \cite{Zhu+23}, the qualitative results remain the same, and the radio AGN population evolves with $z$ from a small $f_{\text{SF}}$ at low redshift to a $f_{\text{SF}}$ value similar to galaxies at high redshift. 

The bottom-left panel of \mbox{Figure \ref{fig 4.2.1}} also reflects higher incidence ratios in its bins. About 50\% of bins have an incidence ratio larger than one, compared to $\approx10\%$ for the previous sample. This new result echos \cite{Igo+24}, suggesting that the incidence of quiescent and star-forming radio AGN are similar and AGNs are not only found in ``red and dead" galaxies. Now, as properties of more bins can be studied due to a larger sample size, we can see a general trend such that AGNs with higher $M_{\star}$ and lower $z$ tend to have lower incidence ratios. Additionally, more regions in the $\log M_{\star}-z-\Delta \text{MS}$ graph can be determined reliably. The positive correlation of $\Delta \text{MS}$ and $z$ and the negative correlation of $\Delta \text{MS}$ and $M_{\star}$ are apparent in the bottom-right panels of \mbox{Figure \ref{fig 4.2.1}}.

We additionally explore the impact of employing radio-samples selected by the $\text{central}-3\sigma$ or $\text{central}-4\sigma$ criteria, as outlined in \mbox{Section \ref{subsec: new AGN selection}}, on our results. The $1\sigma$-selected sample is deemed unsuitable for this investigation due to its low purity. These two selections are more relaxed than the criteria adopted by \citet{Zhu+23} but stricter than the previously utilized $\text{central}-2\sigma$ criterion. The outcomes align with our expectations: the $f_{\text{SF}}$ of radio AGNs gradually increases as we apply looser radio-AGN selection criteria, owing to the increasing completeness and slightly elevated contamination from star-forming galaxies. The $f_{\text{SF}}$ of radio AGNs and $M_{\star}$-matched galaxies is comparable for all the AGN samples at all redshifts.

Our conclusion with the new sample remains similar to \mbox{Section \ref{sec: analysis}}: a significant fraction of radio AGNs are star-forming. Specifically, at $z\gtrsim0.5$, more than half of AGNs are star-forming, and the $f_{\text{SF}}$ is largely comparable to galaxies. For the star-forming AGNs, many of them have higher SFR than that of MS galaxies. This further supports our previous results.  
\begin{figure*}
    \centering
    \includegraphics[width=1.0\textwidth]{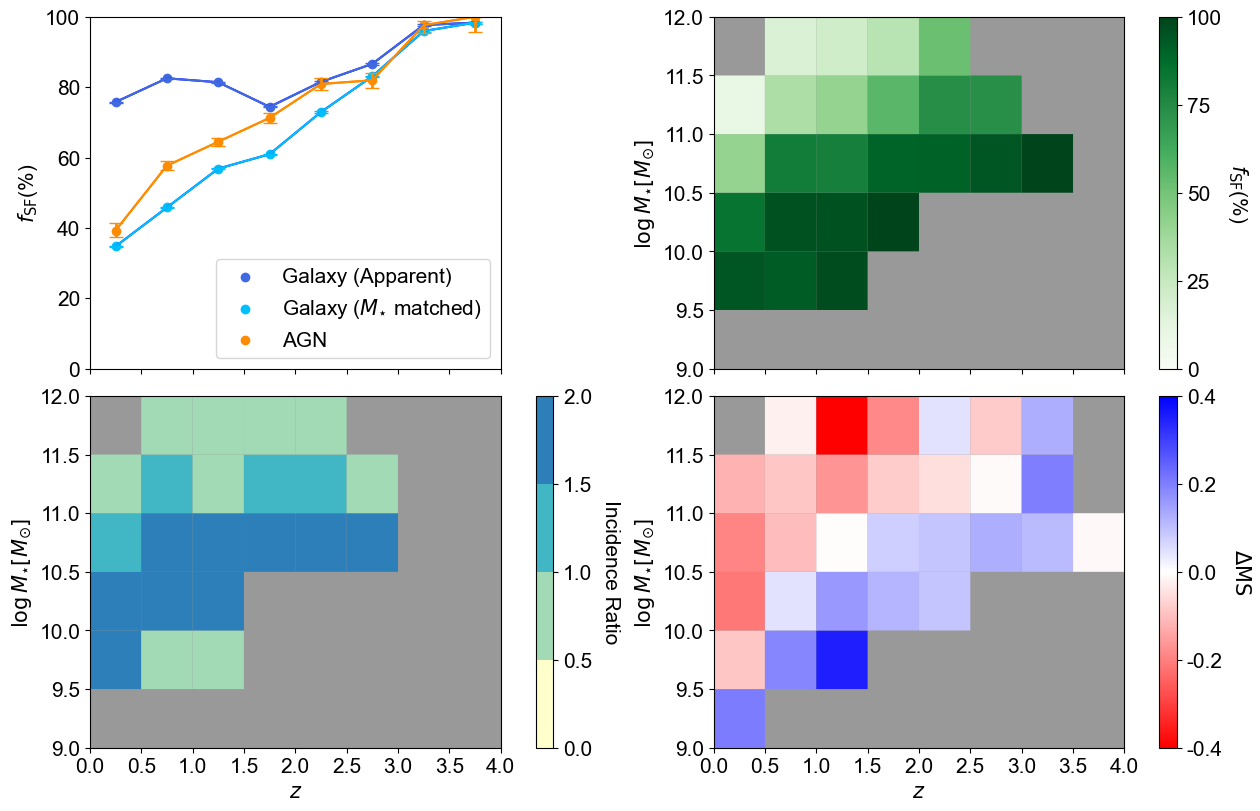}
    \caption{Results for the newly selected radio AGN sample, where the MS is determined with the $UVJ$ method. \textit{Top left}: $f_{\text{SF}}$ of AGNs and galaxies in bins of $z$. \textit{Top right}: $f_{\text{SF}}$ of AGNs in bins of $z$ and $M_{\star}$. \textit{Bottom left}: Ratio of the measured incidence of radio AGNs in star-forming and quiescent galaxies in bins of $z$ and $M_{\star}$. \textit{Bottom right}: $\Delta$MS in bins of $z$ and $M_{\star}$.}
    \label{fig 4.2.1}
    
\end{figure*}

\subsection{Comparison of Radio AGNs Identified and Unidentified in Other Wavebands}  \label{subsec: HERG}
 
Radio AGNs can be classified into high-excitation radio galaxies (HERGs) and low-excitation radio galaxies (LERGs). HERGs and LERGs represent the ``radiative mode" and ``jet mode" of radio AGNs, respectively (e.g., \citealt{Kondapally+22}; \citealt{Magliocchetti+22}). It is generally believed that these two types of radio AGNs have different physical natures. HERGs have more cold gas, which allows for efficient accretion of gas at rates between 1\% and 10\% of the Eddington limit. In contrast, LERGs are primarily fueled by the cooling of hot gas from the galactic halo, with accretion rates below 1\% of the Eddington limit (e.g., \citealt{Allen+06}; \citealt{Heckman+14}).
A small portion ($\approx9\%$) of the radio AGN sample used in this paper has also been identified as X-ray AGNs or MIR AGNs, which are likely to be high-excitation radio galaxies (HERGs), while radio AGNs not identified in other wavebands are mostly low-excitation radio galaxies (LERGs). We can compare these two types of radio AGNs to study the similarities and differences in their star formation properties, thereby gaining insights into the nature of HERGs and LERGs. 

\begin{figure*}
    \centering
    \includegraphics[width=1.0\textwidth]{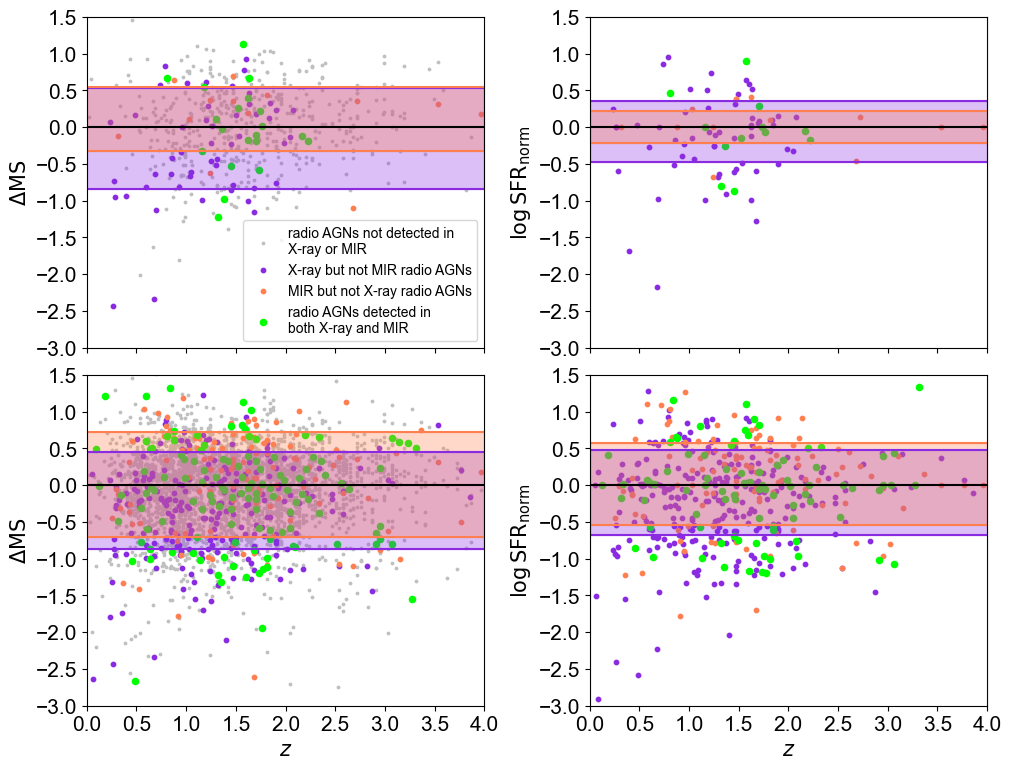}
    \caption{\textit{Top left}: The relation between $\Delta\mathrm{MS}$ and $z$ for different classes of radio AGNs from the \citet{Zhu+23} sample. Silver, blue-violet, red, and green dots represent radio AGNs not identified in other bands, identified in \mbox{X-rays} but not MIR, identified in MIR but not \mbox{X-rays}, and identified in both \mbox{X-rays} and MIR, respectively. The blue-violet and coral shaded regions indicate the $1\sigma$ normalized median absolute deviations from the median $\Delta\mathrm{MS}$ for X-ray-identified and MIR-identified radio AGNs, respectively. \textit{Top right}: Similar to the top-left panel, but depicting $\log\mathrm{SFR}_{\mathrm{norm}}$. The bottom panels are analogous to the top panels but present results from the new sample in this work.}
    \label{fig 4.3}
\end{figure*}

From the 1613 mass-complete radio AGNs selected by \cite{Zhu+23}, 140 are also identified as \mbox{X-ray} AGNs by \cite{Zou+22}, and 35 are classified as MIR AGNs based on the criterion from \cite{Donley+12}. 22 radio AGNs are identified as both \mbox{X-ray} and MIR AGNs. These radio AGNs with other AGN signatures are fitted mainly by star-forming + AGN templates, contrary to those only identified in radio (\citealt{Zou+22}). In the top-left panel of \mbox{Figure \ref{fig 4.3}}, we present the relation between $\Delta\text{MS}$ and $z$ for radio AGNs identified in \mbox{X-rays} and MIR, along with those not identified in other bands, which constitute the majority of the radio AGN population. Although the sample size is limited, we observe that the AGNs identified in other wavebands lie close to the MS. We also show the $1\sigma$ scatter from the median $\Delta\text{MS}$ for these AGNs. We then match X-ray/MIR radio AGNs with all radio AGNs within 0.1 dex of $M_{\star}$ and within $0.075\times(1+z)$ of $z$, similar to \mbox{Section \ref{subsec: deltams}}. For each AGN in the sample, we calculate $\text{SFR}_{\text{norm}}$ as the ratio of its SFR to the median SFR of the matched radio AGNs and show $\log\text{SFR}_{\text{norm}}$ in the top-right panel of \mbox{Figure \ref{fig 4.3}}. We also show the $1\sigma$ scatter from the median $\log\text{SFR}_{\text{norm}}$ for these AGNs. The $\log\text{SFR}_{\text{norm}}$ is close to zero, supporting the conclusion that these AGNs have similar SFRs to the general population of radio AGNs. 

Similarly, we show the results based on the new radio AGN sample in \mbox{Section \ref{subsec: new AGN selection}} in the bottom panels of \mbox{Figure \ref{fig 4.3}}. In this sample, 485 and 247 radio AGNs are identified in \mbox{X-rays} and MIR, respectively, with 129 radio AGNs identified in both \mbox{X-rays} and MIR. The global median $\Delta\text{MS}$ of \mbox{X-ray-identified} radio AGNs ($-0.27\pm0.04$) is lower than that of \mbox{X-ray-unidentified} radio AGNs ($-0.05\pm0.01$), suggesting that \mbox{X-ray-identified} radio AGNs experience more quenching of star formation, with a $\approx0.2$ dex lower SFR. On the other hand, \mbox{MIR-identified} radio AGNs have a similar or slightly higher global mean $\Delta\text{MS}$ ($-0.02\pm0.06$) compared to \mbox{MIR-unidentified} radio AGNs ($-0.07\pm0.01$). When we compare the SFR of radio AGNs identified in other wavebands to that of matched radio AGNs with similar $M_{\star}$ and $z$, we find a global median $\log\mathrm{SFR}_{\mathrm{norm}}$ of $-0.12\pm0.04$ and $0.04\pm0.04$ for \mbox{X-ray-} and MIR-identified radio AGNs, respectively. Our results imply that \mbox{X-ray}-identified radio AGNs exhibit slightly more quenched star formation compared to those not identified in \mbox{X-rays}, while MIR-identified radio AGNs tend to have similar SFRs relative to those not identified in MIR. The divergence is generally small for these two types of AGNs and near zero in the case of MIR-identified and MIR-unidentified radio AGNs. 

As the \mbox{X-ray} and MIR AGNs are more likely to be HERGs and the radio AGNs not identified in other bands are mostly LERGs, our results imply that HERGs and LERGs have little difference in their SFRs. This conclusion agrees with \cite{Zhu+23} and is generally consistent with that of \cite{Whittam+12} that the host-galaxy SFRs and $M_{\star}$ of HERGs and LERGs are similar at high redshifts.


\section{Summary and Future Work} \label{sec: conclusion}

In this work, we present a comprehensive study of the star-forming characteristics of radio AGNs using a large and complete sample of galaxies and AGNs from the W-CDF-S, ELAIS-S1, and XMM-LSS fields. Leveraging the rich multiwavelength data available in these fields, we employed two different methods to select radio AGN samples: the strict criterion based on radio excess from \cite{Zhu+23} and a new selection using the IRRC and its dependence on $M_{\star}$ and $z$. We also defined star-forming galaxies and the MS using two independent methods based on $UVJ$ colors and nSFRs. 

Our main findings can be summarized as follows:

\begin{enumerate}
    \item The $f_{\text{SF}}$ of radio AGNs exhibits a strong positive evolution with $z$ and a negative trend with $M_{\star}$, broadly mirroring the behavior of galaxies. For our main radio AGN sample, at low redshifts ($z < 0.5$), only $\approx10\%$ of radio AGNs are star-forming, much smaller than that of galaxies ($\approx75\%$). However, by $z \approx 2-2.5$, the $f_{\text{SF}}$ of radio AGNs increases to $\approx65\%$, approaching the values seen in the general galaxy population. See \mbox{Section \ref{subsec: frac}}.

    \item After matching radio AGNs to a control sample of galaxies with similar $M_{\star}$ and $z$, we find that the tendency of radio AGNs to reside in massive galaxies is the primary driver of their apparent low $f_{\text{SF}}$. Once the host-galaxy properties are accounted for, the $f_{\text{SF}}$ of radio AGNs becomes nearly comparable to that of galaxies at all redshifts. See \mbox{Section \ref{subsec: frac}}.

    \item For the star-forming subset of radio AGNs, we investigated their position relative to the MS by computing $\Delta \text{MS}$. We find a positive correlation between $\Delta \text{MS}$ and $z$ and a negative correlation with $M_{\star}$. Radio AGNs in massive, low-redshift galaxies tend to have lower SFRs than the MS. At the same time, those in low-mass, high-redshift systems exhibit comparable or even enhanced star-formation activity. See \mbox{Section \ref{subsec: deltams}}.

    \item After applying 1.4 GHz flux limits for the radio-AGN sample to address the selection effect from the correlation between the SFR and the $24\:\mu\text{m}$ flux, we investigated whether the radio luminosity of radio AGNs influences their host-galaxy star-formation characteristics. We found that $L_{1.4\text{GHz}}$ has only minor influence on $f_{\text{SF}}$ or $\Delta\text{MS}$. See \mbox{Section \ref{subsec: different Lradio}}.

    \item Our results remain qualitatively consistent when using different definitions of the MS ($UVJ$ vs. nSFR) and various radio AGN selection criteria, although the specific values of $f_{\text{SF}}$ and $\Delta \text{MS}$ can vary. The larger radio AGN sample selected via the IRRC shows a higher overall $f_{\text{SF}}$, but the general trends with $z$ and $M_{\star}$ are preserved. See \mbox{Section \ref{subsec: effect of AGN}} and \mbox{Appendix \ref{appen1}}.

    \item Radio AGNs detected in the \mbox{X-ray} and MIR only show small ($\lesssim0.2$ dex) differences in SFR compared to those not detected. This agrees with the finding by \cite{Whittam+12} that the host-galaxy SFRs and $M_{\star}$ of HERGs and LERGs are similar at high redshifts. See \mbox{Section \ref{subsec: HERG}}.
    
\end{enumerate}

These findings indicate that radio AGNs can both suppress and enhance star formation in their host galaxies, depending upon the properties of the galaxy and the cosmic epoch. Radio AGNs are preferentially found in massive, quiescent systems at low redshifts, where their feedback is likely responsible for quenching star formation. However, at higher redshifts ($z \gtrsim 1.5$), a significant fraction of radio AGNs resided in star-forming galaxies, and many of these systems exhibit elevated SFRs compared to MS galaxies of similar $M_{\star}$. 

Our work highlights the importance of constructing complete, multiwavelength samples and carefully accounting for selection effects when studying the star-formation properties of AGN populations. The results underscore the complex interplay between nuclear activity and galaxy evolution, with radio AGNs playing a dual role in regulating and potentially enhancing star formation over cosmic time. 
Looking ahead, ongoing and future deep multiwavelength surveys will enable the construction of even more complete and unbiased samples of radio AGNs and galaxies. Radio surveys like the ongoing MIGHTEE survey (e.g., \citealt{Heywood+22}; \citealt{Gurkan+22}), the ongoing Evolutionary Mapping of Universe (EMU) survey conducted with the Australian Square Kilometre Array Pathfinder (e.g., \citealt{Joseph+19}) and the upcoming Square Kilometer Array (SKA; e.g., \citealt{Dewdney+09}; \citealt{Norris+13}; \citealt{McAlpine+15}) will allow more galaxies and AGNs to be detected in the radio and extensively studied. Upcoming photometric surveys can also greatly improve the quality of photometric redshifts and SEDs for our radio AGNs. For example, the W-CDF-S field will be deeply observed by Euclid in the near-infrared; the upcoming LSST DDF observations will also provide much deeper $ugrizy$ data in the W-CDF-S, ELAIS-S1, and XMM-LSS fields. All of these are opportunities to improve the measurement of host-galaxy properties and to study the star-formation characteristics of radio AGNs. Many spectroscopic surveys are being or will be conducted in our fields, including the Deep Extragalactic Visible Legacy Survey (DEVILS; \citealt{Davies+18}), the Multi-Object Optical and Near-Infrared Spectrograph (MOONS; \citealt{Maiolino+20}) survey, the Subaru Prime Focus Spectrograph (PFS; \citealt{Takada+14}) survey, and the Wide Area VISTA Extragalactic Survey (WAVES; \citealt{Driver+19}). These surveys can provide high-quality redshifts and rich spectroscopic data for source selection and characterization. Specifically, new spectroscopic data can provide much better samples of HERGs and LERGs for exploring their similarities and differences. These potential studies of the star-formation characteristics of radio AGNs will allow for more detailed investigations into the physical drivers governing the coevolution of supermassive black holes and their host galaxies, shedding light on the mechanisms responsible for fueling nuclear activity and shaping the stellar populations of galaxies across cosmic time.

\section*{Acknowledgments}
\begin{acknowledgments}
B.Z. acknowledges support from USTC. F.Z., W.N.B., N.C., and Z.Y. acknowledge support from NSF grants AST-2106990 and AST-2407089, Chandra X-ray Center grant AR4-25008X, and the Penn State Eberly Endowment. S.Z. and Y.X. acknowledge support from NSFC grant 12025303.

\end{acknowledgments}

\appendix
\section{Using the nSFR method to select star-formation galaxies} \label{appen1}

Our main method to select star-forming galaxies is to use a $(U-V)-(V-J)$ color diagram. Different star-forming-galaxy definitions can alter the shape of the derived MS. Thus, to assess the robustness of our results, we use another definition to select a star-forming galaxy sample. The dimensionless normalized SFR (nSFR), calculated as $\text{SFR}\times t_H/M_{\star}$, is the ratio of the current SFR and the time-averaged SFR over the whole star-forming history ($t_{H}$ is the Hubble time at the relevant redshift). nSFR can also be utilized to differentiate the star-forming and non-star-forming populations (e.g., \citealt{Pacifici+16}; \citealt{Carnall+18}; \citealt{Kondapally+22}). Following \cite{Kondapally+22}, we select star-forming galaxies as those satisfying $\text{nSFR}>1/5$. Galaxies selected by this criterion have been shown to agree well with $UVJ$-selected galaxies (\citealt{Williams+09}). Using the nSFR method, we select 2457397 (92.2\%) reference star-forming galaxies for further study.
We apply the same methods to the radio AGNs and select 680 (39.6\%) and 4181 (61.8\%) star-forming ones from the \cite{Zhu+23} radio AGN sample and the $q_{\text{IRRC}}$-selected new sample, respectively. 

\subsection{The MS}

For the nSFR method, we also computed the MS. \mbox{Figure \ref{fig 2.3}} shows the relationship between SFR and $M_{\star}$ at six redshifts ($z=$ 0.5, 1.0, 1.5, 2.0, 2.5, 3.0). The MS determined by $UVJ$- and nSFR-selected star-forming galaxies are both shown in this figure. The two MS generally agree well. However, they diverge by $\lesssim0.5$ dex at $z<1$ for the high-$M_{\star}$ end. The MS for massive galaxies is sensitive to the adopted definition of star-forming galaxies, and the large divergence at low redshifts is caused by the significant number of quiescent and transitioning galaxies in this regime (e.g., \citealt{Donnari+19}; \citealt{Cristello+24}). The nSFR method becomes less reliable at these high redshifts, which causes the $0.2-0.3$ dex divergence at $z\approx2.5-3.0$.

\subsection{Results for \cite{Zhu+23} Radio AGNs}

\begin{figure*}
    \centering
    \includegraphics[width=1.0\textwidth]{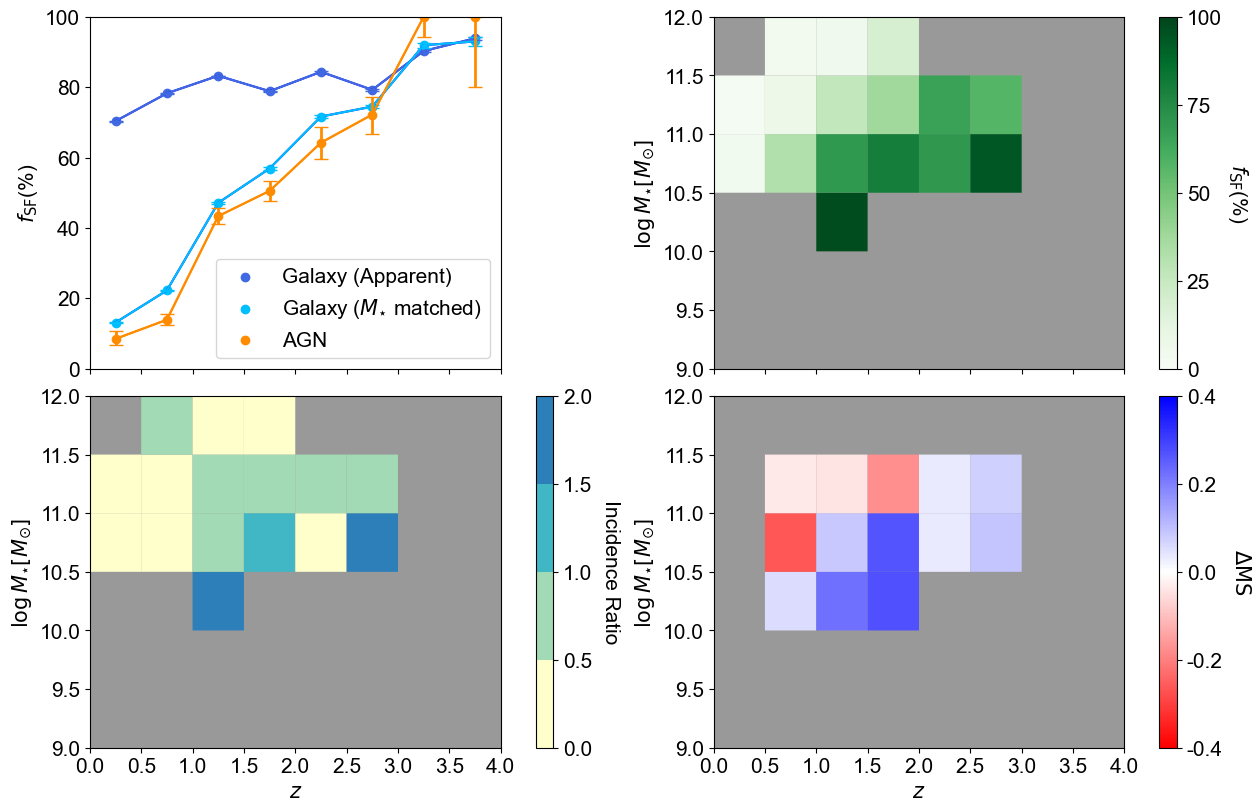}
    \caption{Results for the \cite{Zhu+23} radio AGN sample, where the MS is determined with the nSFR method. The format of the panels is identical to that of \mbox{Figure \ref{fig 4.2.1}}.}
    \label{fig 5.2}
    \end{figure*}

For the MS derived by the nSFR method, we show the results for the \cite{Zhu+23} radio AGN sample in \mbox{Figure \ref{fig 5.2}}. Comparing the top-left and top-right panels with those of \mbox{Figure \ref{fig 3.1}}, we see $f_{\text{SF}}$ of both galaxies and AGNs are $<5\%$ lower than for the $UVJ$ definition in most bins of $z$. From the bottom-left panel, we see that the incidence ratio agrees with that of the $UVJ$ method in most bins, although with a larger difference at high redshifts. The bottom-right panel of $\Delta \text{MS}$ also agrees well with previous results, only with a small difference at the low-$z$ and high-$z$ ends. The AGNs at low redshifts are more quiescent; thus, their positions relative to the MS may vary more across different definitions. The number of AGNs at $z>2.5$ is small, resulting in larger statistical fluctuations. We also consider the ``safe" regime following \cite{Cristello+24}, which is established to minimize the MS offset while probing the highest masses possible. The ``safe" regime is defined as areas with $f_{\text{SF}}$ of galaxies and AGNs larger than 0.5. If we implement the ``safe" regime, a few bins at the high-$M_{\star}$ and low-$z$ regime in the bottom-left and bottom-right panels are considered unreliable for study. This agrees well with our findings when using different MS definitions and does not alter our results.

In conclusion, our results are materially the same under two MS definitions. This further supports our argument in \mbox{Section \ref{subsec: SF selection}} that the colors of radio AGN hosts in our sample seldom suffer from AGN-light contamination. In our case, different MS definitions only influence the $\Delta \text{MS}$ in bins with the most quiescent AGNs or lowest AGN counts. 

\subsection{Results for $q_{\text{IRRC}}$-selected Radio AGNs}

\begin{figure*}
    \centering
    \includegraphics[width=1.0\textwidth]{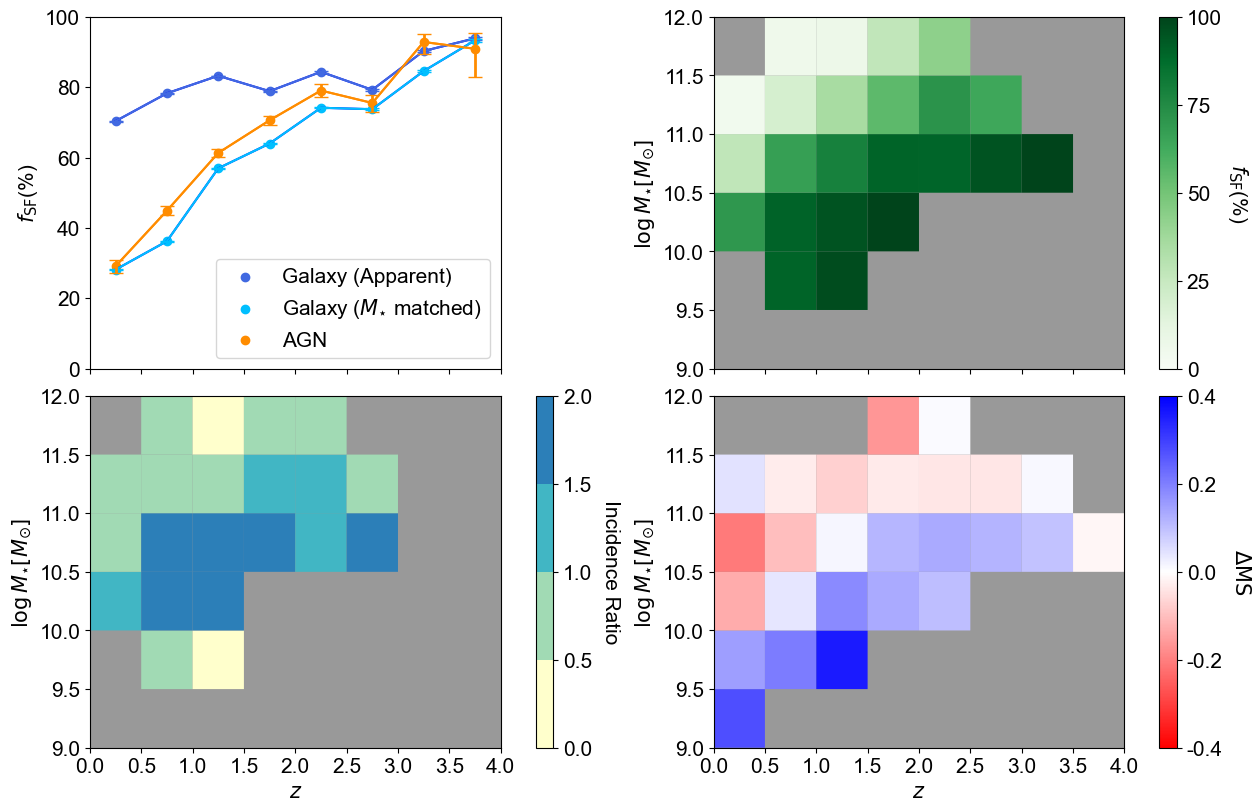}
    \caption{Results for the newly selected radio AGN sample, where the MS is determined with the nSFR method. The format of the panels is identical to that of \mbox{Figure \ref{fig 4.2.1}}.}
    \label{fig 5.3}
    
\end{figure*}

We show the results for $q_{\text{IRRC}}$-selected radio AGNs and the nSFR-derived MS in \mbox{Figure \ref{fig 5.3}}. Results here are also similar to those from the $UVJ$ method. Thus, for both of our radio-AGN samples, different MS definitions almost do not influence our results. 

\bibliography{submit/main.bib}

\end{document}